\documentclass[]{interact}

\usepackage{epstopdf}
\usepackage[caption=false]{subfig}

\usepackage[numbers,sort&compress]{natbib}

\usepackage[latin9]{inputenc}
\usepackage{verbatim}
\usepackage[table]{xcolor}
\usepackage[linesnumbered, ruled, vlined]{algorithm2e}
\usepackage{amsmath, amssymb, amsthm, abstract, amsfonts}
\usepackage{bbm}
\usepackage{color}
\usepackage{epsf, enumerate, epsfig, epstopdf, enumitem, environ}
\usepackage{graphicx}
\usepackage{multirow, multicol, mathrsfs, mathtools}
\usepackage{url}
\usepackage{threeparttable}
\usepackage{xparse}
\makeatletter
\usepackage{natbib}
\usepackage{hyperref}

\bibpunct[, ]{[}{]}{,}{n}{,}{,}
\renewcommand\bibfont{\fontsize{10}{12}\selectfont}

\theoremstyle{plain}
\newtheorem{theorem}{Theorem}[section]

\newtheorem{proposition}[theorem]{Proposition}

\theoremstyle{definition}

\theoremstyle{remark}

\begin{document}

\articletype{Original Research Article}

\title{Rapid Detection of Hot-spots via Tensor Decomposition with Applications to Crime Rate Data}

\author{
\name{
Yujie Zhao\textsuperscript{a},
Hao Yan\textsuperscript{b},
Sarah Holte\textsuperscript{c},
and Yajun Mei\textsuperscript{a}
\thanks{CONTACT Author Email: yzhao471@gatech.edu} }
\affil{
\textsuperscript{a}School of Industrial and Systems Engineering, Georgia Institute of Technology, Atlanta, GA, USA;
\textsuperscript{b}School of Computing, Informatics, and Decision Systems Engineering, Arizona State University, Tempe, AZ, USA;
\textsuperscript{c}Division of Public Health Sciences, Fred Hutchinson Cancer Research Center, Seattle, WA, USA}
}

\maketitle

\begin{abstract}
    We propose an efficient statistical method (denoted as SSR-Tensor) to robustly and quickly detect hot-spots that are sparse and temporal-consistent in a spatial-temporal dataset through the tensor decomposition.
    Our main idea is first to build an SSR model to decompose the tensor data into a Smooth global trend mean, Sparse local hot-spots, and Residuals.
    Next, tensor decomposition is utilized as follows:  bases are introduced to describe within-dimension correlation, and tensor products are used for between-dimension interaction.
    Then, a combination of LASSO and fused LASSO is used to estimate the model parameters, where an efficient recursive estimation procedure is developed based on the large-scale convex optimization, where we first transform the general LASSO optimization into regular LASSO optimization and apply FISTA to solve it with the fastest convergence rate.
    Finally, a CUSUM procedure is applied to detect when and where the hot-spot event occurs.
    We compare the performance of the proposed method in a numerical simulation study and a real-world case study, which contains a dataset including a collection of three types of crime rates for U.S. mainland states during the year 1965-2014.
    In both cases, the proposed SSR-Tensor is able to achieve the fast detection and accurate localization of the hot-spots.
\end{abstract}


\section{Introduction}
    The objective of our research is \textit{hot-spot} detection when monitoring multiple data sources or streams across different spatial regions over time, where one is interested in quickly and accurately determining which data sources or streams change their patterns at which local regions and at which time.
    And in our paper, the definition of hot-spot is the structured outliers that are sparse over the spatial domain but persistent over time.
    A concrete motivating example is to monitor three types of annual crime rates from $1965$ to $2014$ for $48$ mainland states in the United States, see Section \ref{sec: data} below for the detailed data description.
    There are two kinds of changes: one is at the global level, and the other is at the local level.
    For hot-spot detection, we are more interested in detecting those local changes with the following two properties:
    (1) spatial sparsity, i.e., the local changes are sparse in the spatial domain;
    (2) temporal persistence, i.e., the local changes last for a reasonably long time period unless one takes some actions.

    Generally speaking, hot-spot detection in spatio-temporal data can be treated as a change-point detection problem.
    There are three major categories of methodologies and approaches in the literature.
    The first one is the LASSO-based control chart that integrates LASSO estimators for change point detection and declares non-zero components of the LASSO estimators as the hot-spot, see \citep{zou2009multivariate, zou2008LASSO, LassoBased1, vsaltyte2011spatial}.
    Unfortunately, the LASSO-based control chart lacks the ability to separate the local hot-spots from the global trend mean in the spatio-temporal data.
    The second category of methods is the dimension-reduction-based control chart, where one monitors the features from Principal Component Analysis (PCA) or other dimension reduction methods.
    For example, \cite{liu1995PCA} reduces the dimensionality in spatio-temporal data by constructing T2 and Q charts.
    \cite{paynabar2016PCA} combines multivariate functional PCA with change-point models to detect the hot-spots.
    For other dimension reduction methods, please see \citep{paynabar2013PCA, PCA, tensorPCA1, tensorPCA2, bakshi1998PCA} for more details.
    The drawback of PCA or other dimension reduction based methods are the restriction of the change detection problem and the failure to take full advantage of the spatial clustering property of hot-spot.
    The third category of hot-spot detection methods from spatio-temporal data is the decomposition-based method that decomposes the hot-spot from background events.
    For example, \cite{AnomalyInImage} and \cite{SSD} proposed Smooth-Sparse Decomposition (SSD) model for hot-spot detection in the spatio-temporal data.
    SSD can separate hot-spot from the functional mean by utilizing the spatial structure of both the functional mean and hot-spot.
    More references can be found in \citep{zhang2018tensor, yu2019tensor, AnomalyInVideo, yan2014image, li2019tensor}.
    However, these existing approaches investigate structured images or curves data and assume that the hot-spot events are independent over the time domain.

    In this paper, we propose to develop a decomposition-based  hot-spot detection method when the hot-spots are from autoregressive (AR) model, which is typical for time series data. It is worth noting that the spatio-temporal data can often be represented in $3$-dimensional tensor format as "Spatial dimension $\times$ Temporal dimension $\times$ Attributes".
    Our main idea is to decompose this tensor into three components: smooth global trend mean, sparse local hot-spot, and residuals.
    We term our proposed decomposition model as \textit{SSR-Tensor}.
	Furthermore, when fitting the raw data to the SSR-Tensor, we propose to add two penalty functions: one is the LASSO type penalty to guarantee the sparsity of hot-spots, and the other is the fused-LASSO type penalty to the autoregressive properties of a hot-spot or time-series data.
    Through our proposed SSR-Tensor model, we are able to
    (1) detect when the hot-spot happens (i.e., the change point detection problem); and
    (2) localize where and which type of the hot-spots occurs if the change happens (i.e., the spatial localization problem).
    We call the first capacity as \textit{hot-spot detection} and the second capacity as \textit{hot-spot localization}.

    Considerable research has been done on modeling and prediction of the spatio-temporal data.
    Some popular time series models are AR, MA, ARMA model, etc.
    And the parameters can be estimated by the Yule-Walker method \citep{hannan1979determination}, maximum likelihood estimation or the least squares method \citep{hamilton1994time}.
    In addition, spatial statistics have also been extensively investigated in their own right, see  \citep{early_defination_neighbor, ecology, lan2004landslide, elhorst2014spatial}.
    Spatio-temporal models are proposed by combining the time series models with spatial statistics \citep{ZhuJun,lai2015asymptotically}.
    Please see the textbook \citep{ST-model-book} for additional literature and detailed discussion.
    We emphasize that our proposed SSR-Tensor model is different from these existing spatio-temporal models in the sense that its primary objective is for hot-spot detection, not for estimation or prediction.

    While our paper focuses only on a $3$-dimensional tensor due to our motivating application in crime rates, our proposed SSR-Tensor model can easily be extended to any $d$-dimensional tensor ( $d \geq 3$ ). If we have additional dimensions, such as the unemployment rate and economic performance, these can be modeled as the additional dimension in the tensor analysis.
    The reason is that our proposed model uses the \textit{basis} to describe correlation within each dimension, and utilizes \textit{tensor products} for interaction between different dimensions.
    Thus, as the dimension  $d$ increases, we just need to add the corresponding bases.
    The capability of extending to high-dimensional data is the main advantage of our proposed SSR-Tensor model.

    The remainder of this paper is as follows:
    Section \ref{sec: data} introduces and visualizes the crime rate dataset, which is used as our motivating example.
    Section \ref{sec: methodology} presents our proposed SSR-Tensor model and discusses how to estimate model parameters from data.
    Section \ref{sec: Temporal Detection and Spatial Location of Hot Spot} describes how to use our proposed SSR-Tensor model to find hot-spots, both for detection and localization.
    Our proposed methods are validated through extensive simulations in Section \ref{sec: simulation} and the case study using the crime rate dataset is shown in Section \ref{sec: case study}.

\section{Motivating Example \& Background}
\label{sec: data}
    This section gives a detailed description of the crime rate dataset used in this paper. The dataset is available from the website of the U.S. Department of Justice Federal Bureau of Investigation, see \url{ https://www.ucrdatatool.gov/Search/Crime/State/StateCrime.cfm}.
    The crime rate dataset is recorded from 1965 to 2014 for 48 mainland states in the U.S. annually.
    In each year and for each state, three types of crime crates are recorded:
    a) \textit{Murder and non-negligent manslaughter};
    b) \textit{Legacy rape}; and
    c) \textit{Revised rape}.
    These three annual crime rates are denoted by $r_{1}$, $r_{2}$ and $r_{3}$, respectively.

\begin{figure}[htbp]
   \centering
   \begin{tabular}{cc}
      \includegraphics[width=0.45\textwidth]{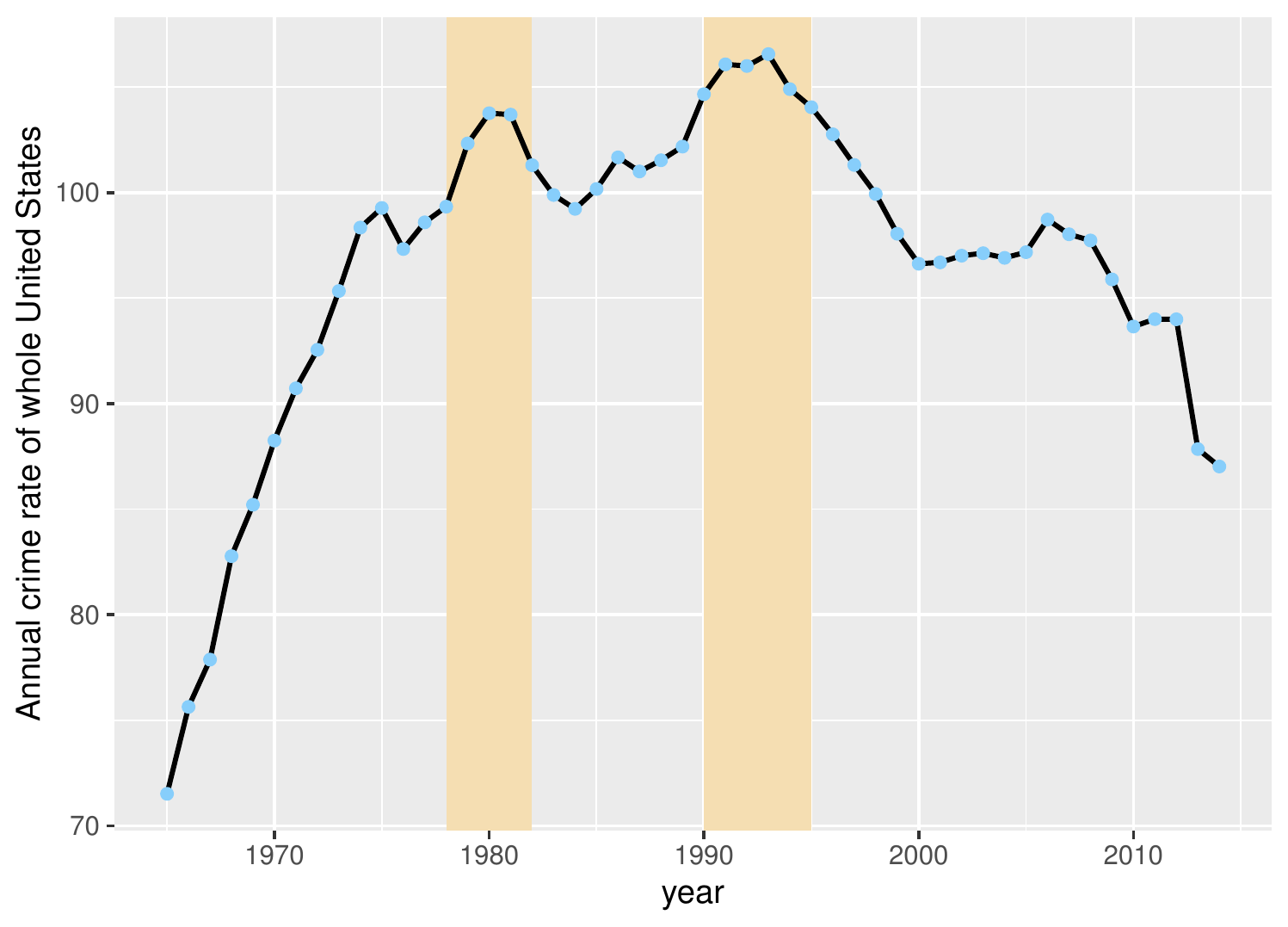} &
      \includegraphics[width=0.45\textwidth]{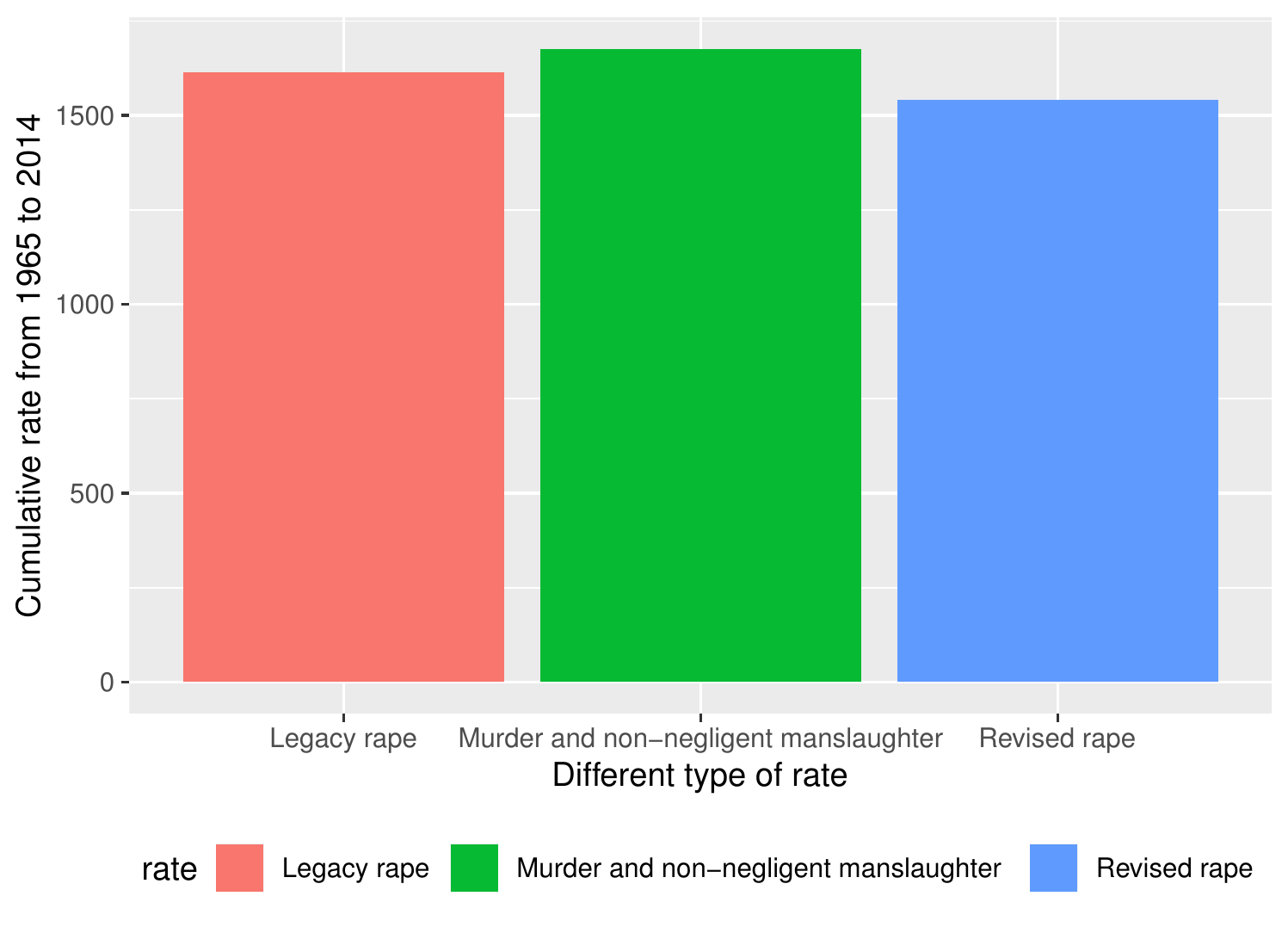} \\
      (a) Time Series plot &
      (b) Bar plot
   \end{tabular}
   \caption{Time Series of Annual Crime Rates in the US over the $50$ years during 1965-2014 (left) \&
            Bar plot of three cumulative rates from 1965 to 2014 (right)
   \label{fig: data description of time and rate domain} }
\end{figure}

    It is worth noting that our motivating dataset is of three dimensions, which includes the year (temporal dimension), state (spatial dimension) and different types of crime rates (attribute/category dimension).
    For the purpose of clarity and visual representation, we plot several figures which shows the characteristic of each dimension.
    To begin with, we first show the characteristic of the temporal domain (year), where we plot the time series of the annual crime rate for the entire U.S. in Figure \ref{fig: data description of time and rate domain}(a).
    The x-axis is the year ranging from 1965 to 2014, and the y-axis is the annual crime rate of U.S..
    It can be seen that the crime rates are increasing in the first ten years during 1965-1975, then become stationary during 1975-1995, and finally have a decreasing trend during 1995-2014.
    Furthermore, we also highlight the two peaks around 1980 and 1992 as well, because the reason to cause these two peaks fails to be determined easily. Probably it is caused by the global trend, or it is also possible to be caused by the local hot-spots.


    Then we show the characteristic of our motivating data on the category domain (type of the crime rates) in Figure \ref{fig: data description of time and rate domain}(b), where different bars represents different type of the crime rates, and the height of the bar represents the cumulative crime rate from 1965 to 2014.
    It can be seen that, these three crime rates overall happen with similar frequencies, which makes it different to detect the hot-spot if we compress the three-dimension data into one dimension only.

\begin{figure}[htbp]
   \centering
   \begin{tabular}{ccc}
      \includegraphics[width=0.31\textwidth]{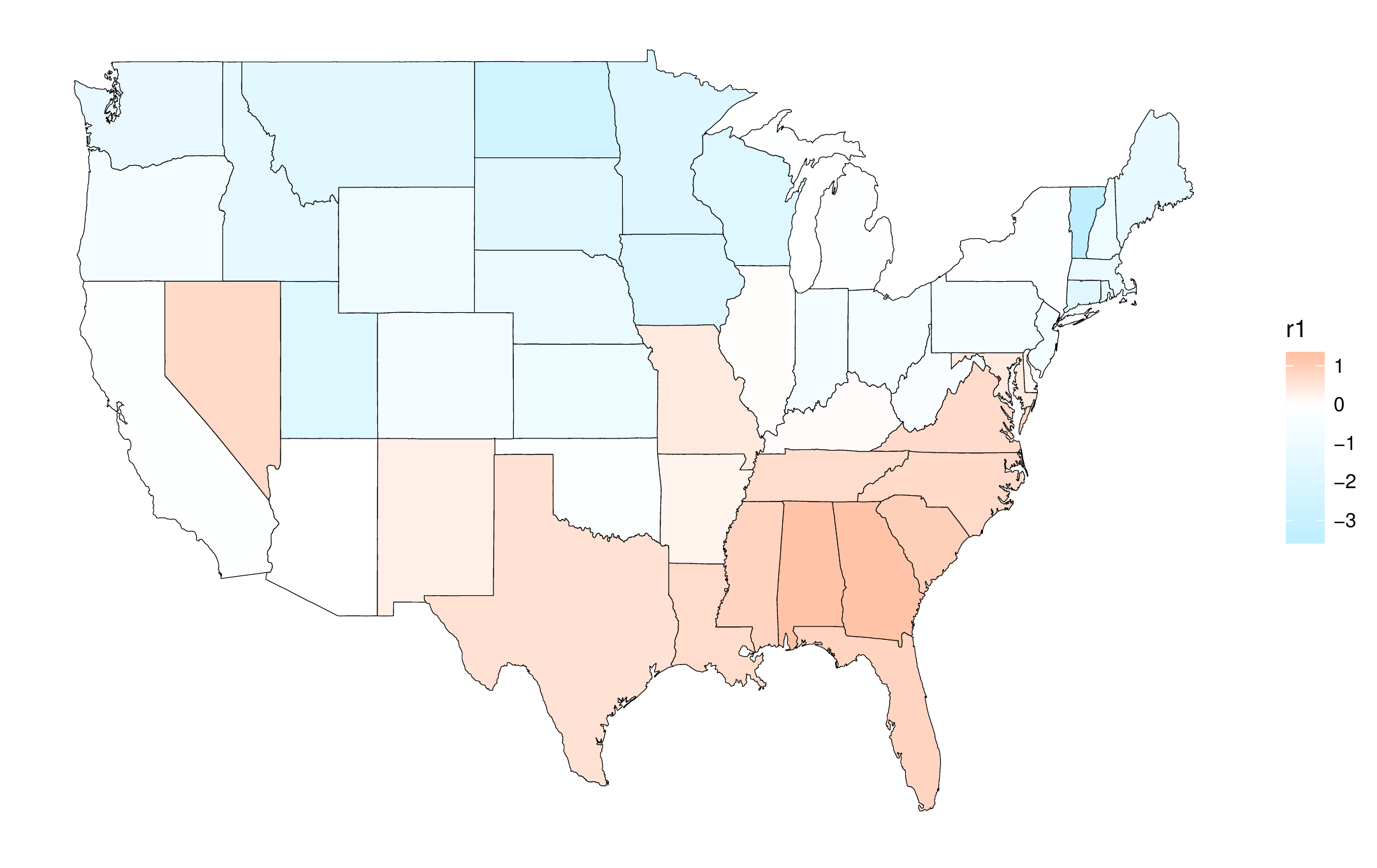} &
      \includegraphics[width=0.31\textwidth]{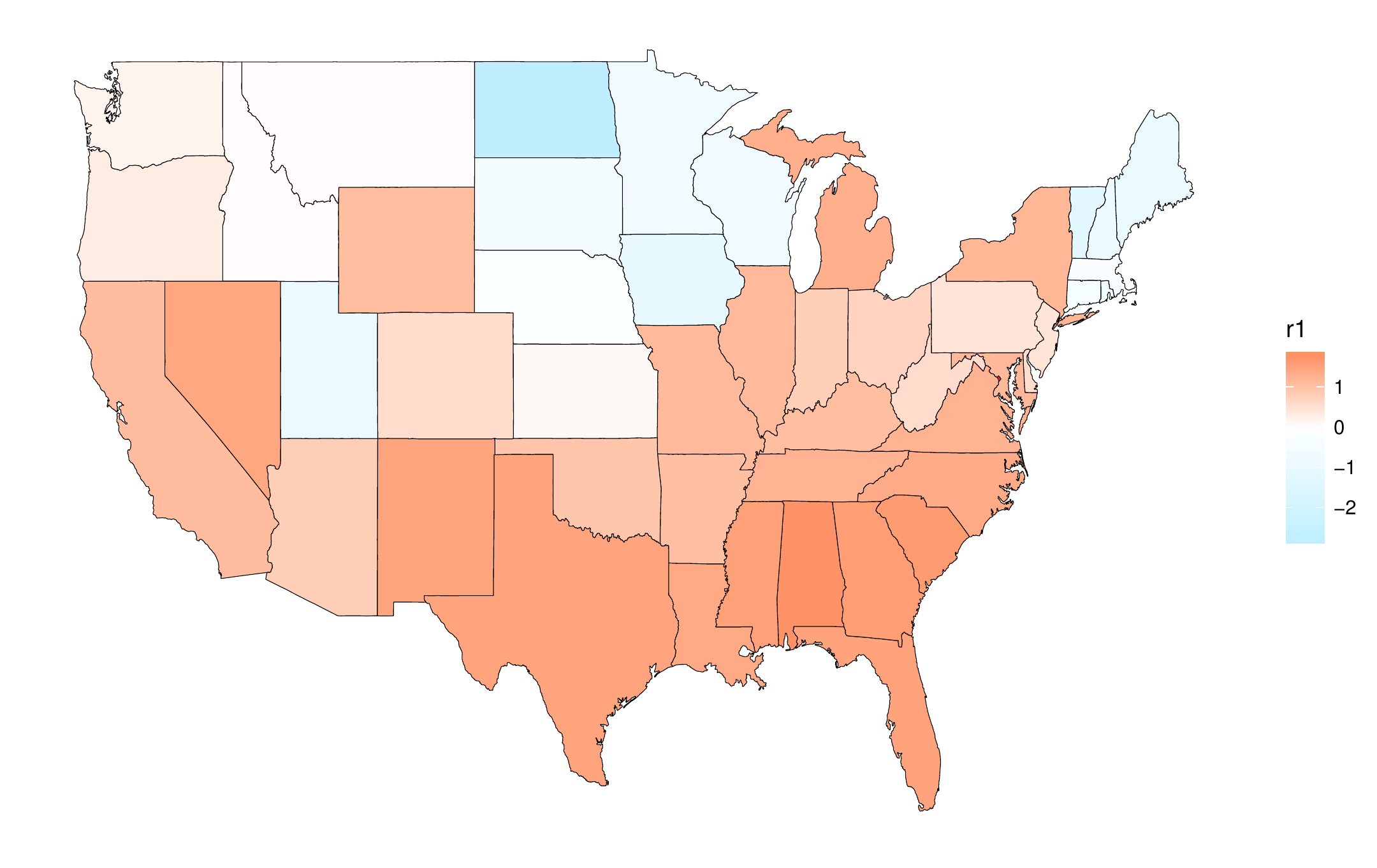} &
      \includegraphics[width=0.31\textwidth]{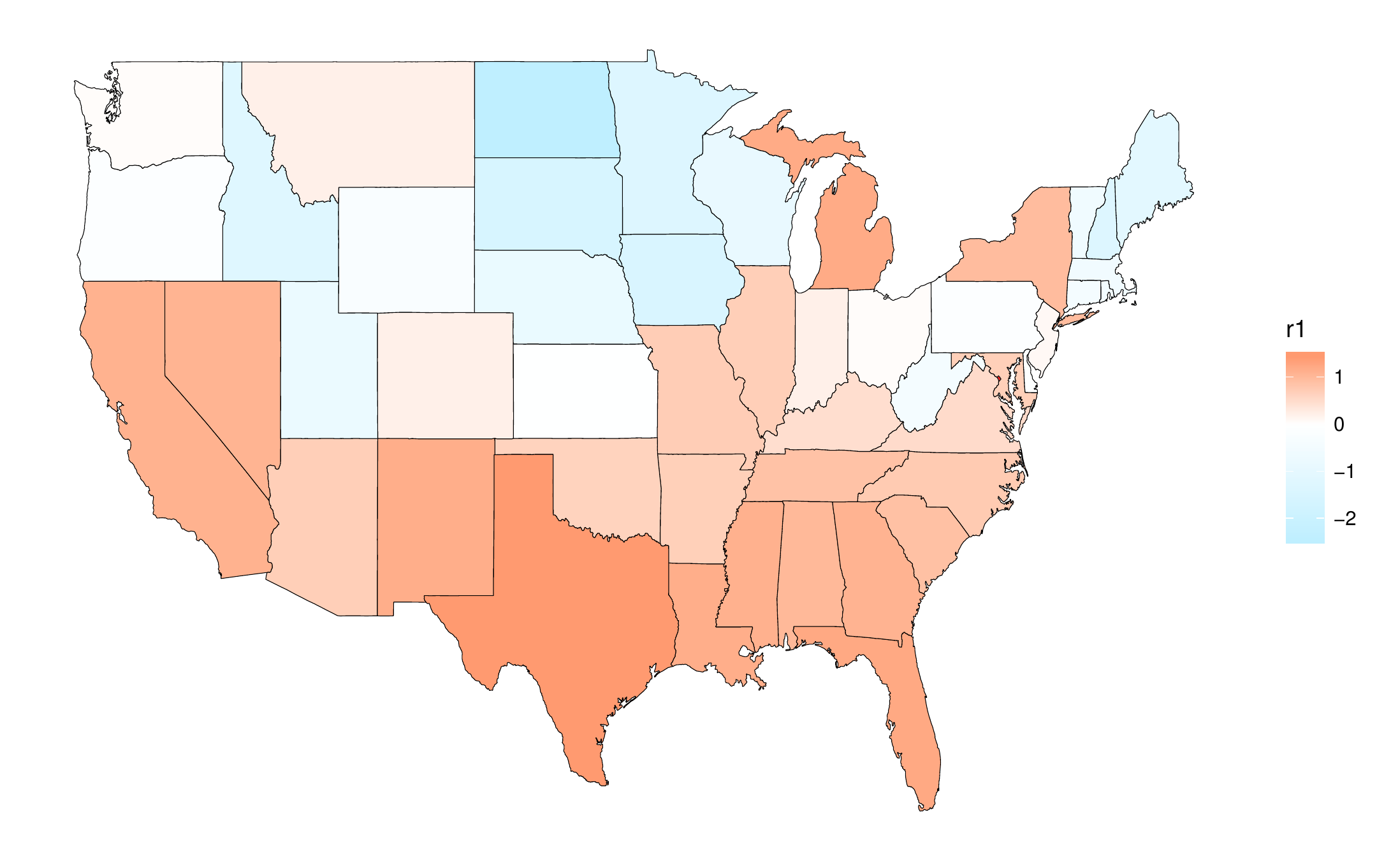}\\
      (a) 1965 &
      (b) 1975 &
      (c) 1985 \\
      \includegraphics[width=0.31\textwidth]{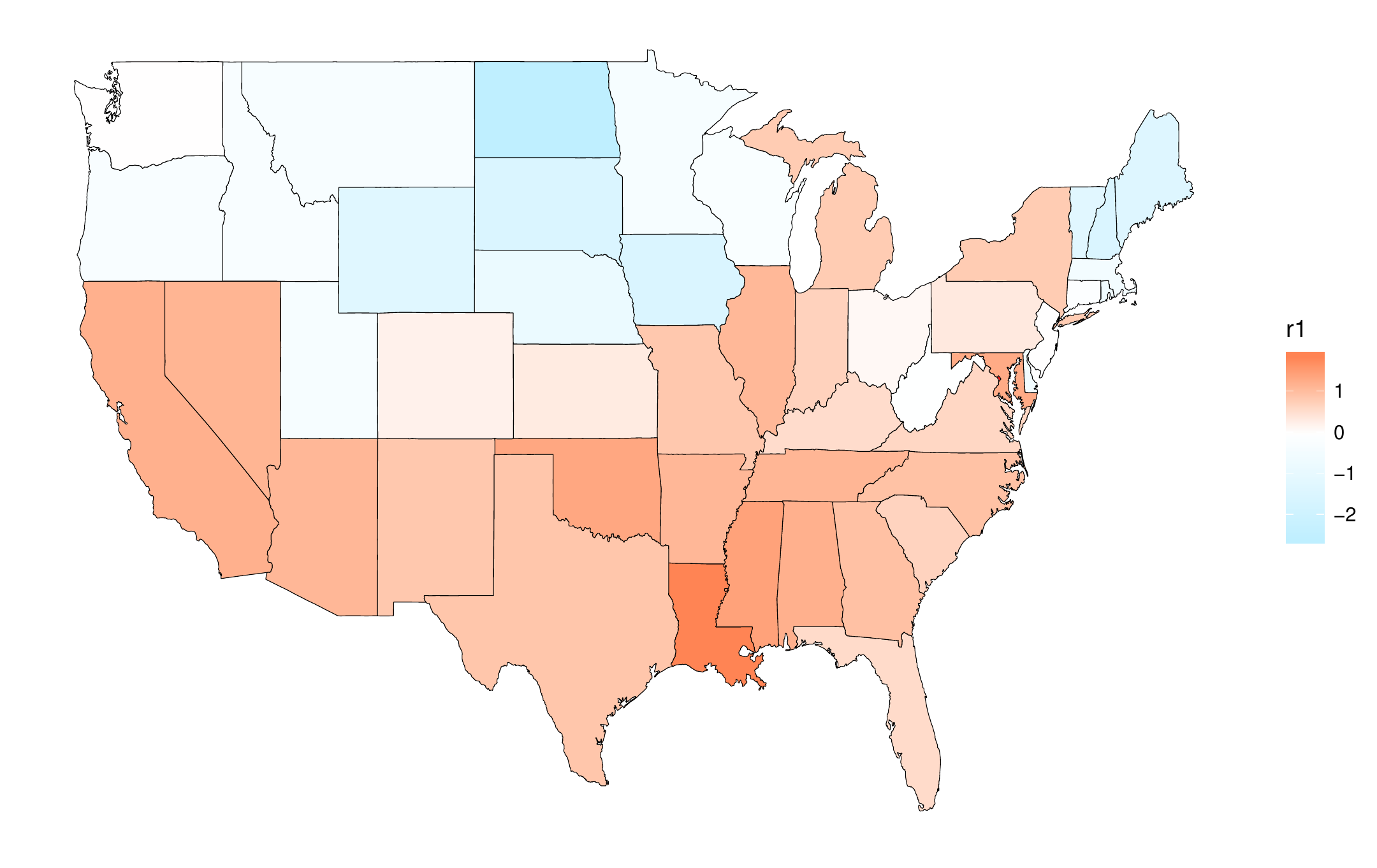} &
      \includegraphics[width=0.31\textwidth]{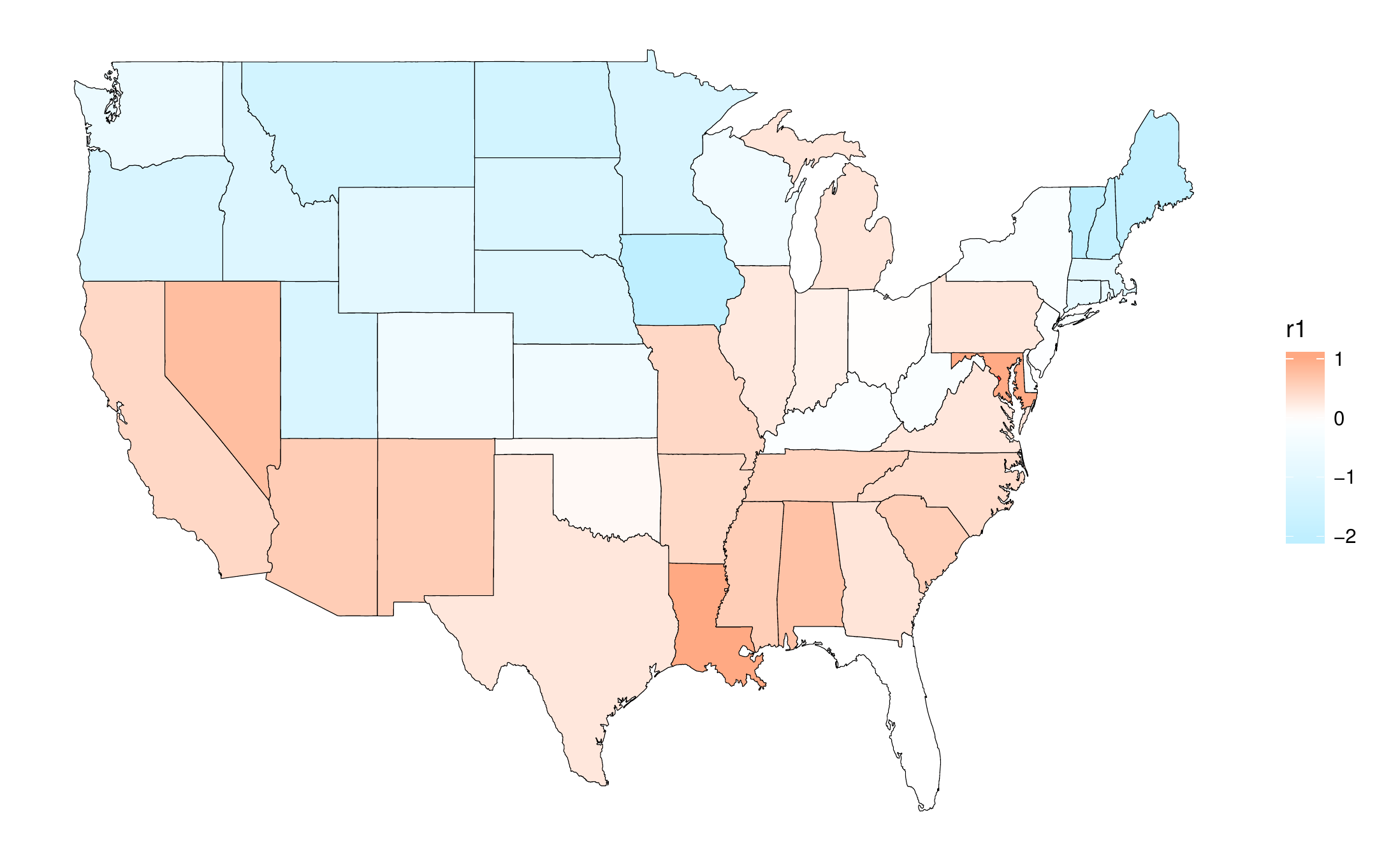} &
      \includegraphics[width=0.31\textwidth]{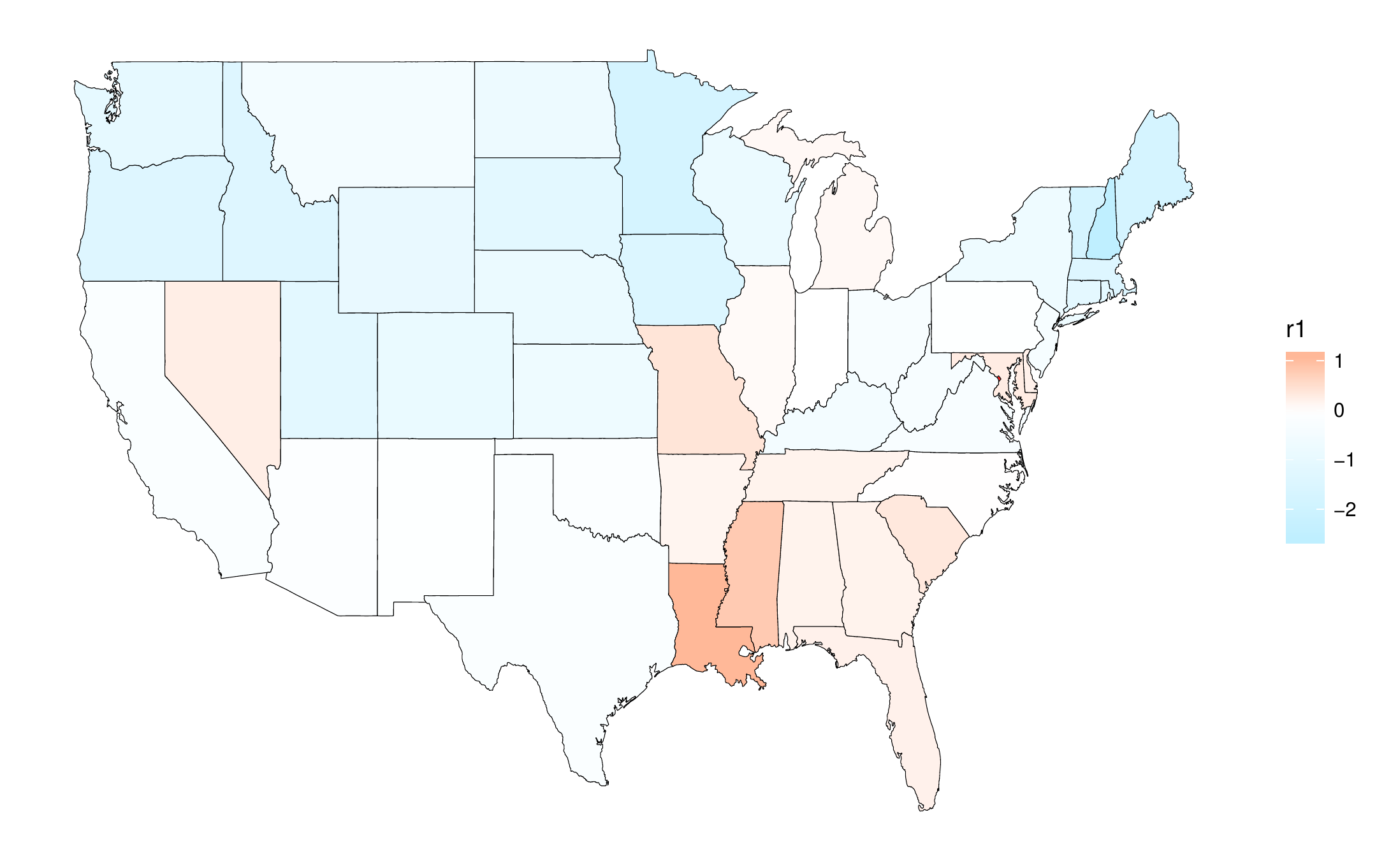} \\
      (d) 1995 &
      (e) 2005 &
      (f) 2014 \\
   \end{tabular}
   \caption{Raw Data for the the third type of crime rate ($r_{3}$) in six different years. From left to right, the  top row represents Year $1965$, 1975 and $1985$, whereas the bottom row are from Year $1995$, $2005$ and $2014$. The red means high crime rates, whereas the blue implies low crime rates.
   \label{fig: Raw Data map}}
\end{figure}

    Finally, we show the characteristic of our motivating data on the spatial domain (state) in Figure \ref{fig: Raw Data map}.
    In Figure \ref{fig: Raw Data map}, each map shows the spatial information  of the third crime rate (revised rape) in six different years.
    And the selected six years is starting from 1965, and ranging with a ten-year interval.
    For the sixth map, since the data in Year 2015 is not available yet, we use Year 2014 instead.
    We can see from the spatial plot in Figure \ref{fig: Raw Data map} that the crime rate spatial patterns are different in different years.
    In addition, the crime rates are increasing in the first ten years and are  decreasing in the last ten years, which is consistent with that in Figure \ref{fig: data description of time and rate domain}(a).

    From Figure \ref{fig: data description of time and rate domain} and Figure \ref{fig: Raw Data map}, there seems to have a brief increasing trend during $1984$-$1995$, but it is difficult to visually conclude  whether this is due to the global change or local hot-spots without refined analysis.
    If this is a local hot-spot event, we want to detect when the hot-spot appears, and then identify where and which type of crime rate accounts for this change.
    Since the data have three dimensions (states, rates and year), how to properly model the global trend mean and local hot-spot structures become very crucial, which is the motivation for our proposed model described in Section \ref{sec: model}.

    It is worth noting that we store our data as a three-dimension tensor, and we noted it as $\mathcal{Y}$.
    Mathematically speaking, the element $\mathcal{Y}_{i,j,t}$ represents the $j$-th crime rate of state $i$ in year $t$, where $i=1,\ldots, 48$ for 48 mainland states, $j=1,\ldots, 3$ for three different type of crime rate and $t = 1, \ldots, 50$ for 50 years from 1965 to 2014.
    In the next sections, we will model this multi-dimensional array through tensor decomposition.
    For the convenience of notation and ease of understanding, we first introduce some basic tensor algebra and notation, including basic notation, definitions, and operators in tensor (multi-linear) algebra that are useful in this paper.
    Throughout the paper, 
    scalars are denoted by lowercase letters (e.g., $\theta$), 
    vectors are denoted by lowercase boldface letters ($\boldsymbol{\theta}$),
    matrices are denoted by uppercase boldface letter ($\boldsymbol{\Theta}$),
    and tensors by curlicue letter ($\vartheta$).
    For example, an order-$K$ tensor is represented by
    $
    \vartheta \in \mathbb{R}^{I_{1} \times \cdots \times I_{K}}
    $,
    where $I_{k}$ represent the mode-$n$ dimension of $\vartheta$ for $k = 1, \ldots, K$.
    The mode-$n$ product of a tensor
    $
    \vartheta \in \mathbb{R}^{I_{1} \times \ldots \times I_{N}}
    $
    by a matrix $\mathbf{B} \in \mathbb{R}^{J_{n}\times I_{n}}$ is a tensor
    $
    \mathcal{A} \in \mathbb{R}^{ I_{1} \times \ldots I_{n-1} \times J_n \times I_{n+1} \times \ldots I_{N} }
    $,
    denoted as
    $
    \mathcal{A} = \vartheta \times_n \mathbf{B},
    $
    where each entry of $\mathcal{A}$ is defined as the sum of products of corresponding entries in $\mathcal{A}$ and $\mathbf{B}$:
    $
    \mathcal{A}_{i_1,\ldots, i_{n-1},j_{n},i_{n+1}, \ldots, i_N}
    =
    \sum_{i_{n}}
    \vartheta_{i_1, \ldots, i_{N}}
    \mathbf{B}_{j_n,i_n}
    $.
    Here we use the notation $\mathbf{B}_{j_n,i_n}$ to refer the $(j_n, i_n)$-th entry in matrix $\mathbf{B}$,
    And the notation $\vartheta_{i_1, \ldots, i_{N}}$ is used to refer to the entry in tensor $\vartheta$ with index $(i_1, \ldots, i_{N})$.
    And the notation
    $
    \mathcal{A}_{i_1,\ldots, i_{n-1},j_{n},i_{n+1}, \ldots, i_N}
    $
    to refer the entry in tensor
    $\mathcal{A}$ with index
    $(i_1,\ldots, i_{n-1},j_{n},i_{n+1}, \ldots, i_N)$.
    
    The mode-n unfold of the tensor $\vartheta \in \mathbb{R}^{I_{1} \times \ldots \times I_{N}}$ is denoted by
    $
    \vartheta_{(n)} \in \mathbb{R}^{I_n \times (I_1\times \ldots I_{n-1} \times I_{n+1} \times I_N)},
    $
    where the column vector of $\vartheta_{(n)}$ are the mode-n vector of $\vartheta$.
    And the mode-n vector of $\vartheta$ are defined as the $I_n$ dimensional vector obtained from $\vartheta$ by varying the index $i_n$ while keeping all the other indices fixed.
    For example, $\vartheta_{:,2,3}$ is a model-1 vector.
    
    A very useful technique in tensor algebra is Tucker decomposition, which decomposes a tensor into a core tensor multiplied by a matrix along each mode:  $\mathcal{Y}=\vartheta\times_{1}\mathbf{B}^{(1)}\times_{2}\mathbf{B}^{(2)}\cdots\times_{K}\mathbf{B}^{(K)}$, where $\mathbf{B}^{(k)}$ is an orthogonal $I_{k}\times I_{k}$ matrix and is a principal component mode-$k$.
    Tensor product can be represented equivalently by a Kronecker product, i.e.,
    $
    \mathrm{vec}(\mathcal{Y})
    =
    (\mathbf{B}^{(K)} \otimes \cdots \otimes \mathbf{B}^{(1)}) \mathrm{vec} (\boldsymbol{\theta})
    $,
    where $\mathrm{vec}(\cdot)$ is the vectorized operator defined as
    $
    \mathrm{vec}(\mathcal{Y})
    =
    \mathcal{Y}_{(K+1)}
    $
    (an $I_{1}\times I_{2}\times\cdots\times I_{K}$-dimension vector).
    The definition of Kronecker product is as follow: Suppose $\mathbf{B}_{1}\in\mathbb{R}^{m \times n}$ and $\mathbf{B}_{2}\in\mathbb{R}^{p\times q}$ are matrices, the Kronecker
    product of these matrices, denoted by $\mathbf{B}_{1}\otimes\mathbf{B}_{2}$, is an $mq\times nq$ block matrix defined by
    $$
    \mathbf{B}_{1}\otimes\mathbf{B}_{2}
    =
    \left[\begin{array}{ccc}
    b_{11}\mathbf{B}_2 & \cdots & b_{1n}\mathbf{B}_2 \\
    \vdots             & \ddots & \vdots             \\
    b_{m1}\mathbf{B}_2 & \cdots & b_{mn}\mathbf{B}_2
    \end{array}\right].
    $$

\section{Our Proposed SSR-Tensor Model}
\label{sec: methodology}
    This section presents our proposed methodology.
    Since the crime rate data is of three dimensions, namely states, rates and year, it will likely have complex within-dimension and between-dimension relationships.
    A within-dimension relationship includes within-state correlation, within-crime-type correlation, and within-year correlation.
    Between-dimension correlations include between-state-and-crime-type interaction, between-state-and-year interaction, as well as between-year-and-crime-type interaction.
    In order to handle these complex ``within-dimension'' and ``between-dimension'' interaction structures, we  use the tensor decomposition method, where basis is used to address ``within'' correlation, and the tensor product is used for ``between'' interaction.
   The choice of basis is also important since different basis can represent different spatial or temporal patterns.
    The detailed discussion on the selection of basis is developed in Section \ref{sec: choice of basis}.

    The structure of this section is as follows:
    Section \ref{sec: model} presents our proposed model that is able to characterize the complex correlation structures;
    Section \ref{sec: Optimization Algorithm} develops the optimization algorithm to solve the estimation problem in Section \ref{sec: model};
    Section \ref{sec: choice of basis} discusses the choice of basis in practice.

\subsection{Our Proposed Model}
\label{sec: model}

    We will use order-three tensor as an example to develop the methodology, as it applies to the crime rate dataset which has three components and can be represented as a three-dimension tensor $\mathcal{Y}_{n_{1}\times n_{2}\times T}$ with $n_1=48$ mainland states, $n_2=3$ different types of crime rates, and $T=50$ years.

    Note that the $i^{th}$,$j^{th}$, and $k^{th}$ slice of the 3-D tensor along the dimension of state, crime type, and year can be achieved as $\mathcal{Y}_{i::},\mathcal{Y}_{:j:},\mathcal{Y}_{::k}$ correspondingly, where $i = 1\ldots n_{1}$, $j = 1\ldots n_{2}$ and $ k = 1 \ldots T$.
    For simplicity, we denote $\mathbf{Y}_{k} = \mathcal{Y}_{::k}$.
    We further denote $\mathbf{y}_{k}$ as the vectorized form of $\mathbf{Y}_{k}$, and $\mathbf{y}$ as the vectorized form of $\mathcal{Y}$.

    The key idea of our proposed model is to separate the global trend mean from the local pattern by decomposing the tensor $\mathbf{y}$ into three parts, namely the smooth global trend mean $\boldsymbol{\mu}$, local hot-spots $\mathbf{h}$, and residuals $\mathbf{e}$, i.e., $\mathbf{y}=\boldsymbol{\mu}+\mathbf{h}+\mathbf{e}$.
    For the first two of the components (e.g., the global time trend mean and local hot-spots), we introduce a basis decomposition framework to represent the structure of the within correlation in the global background and local hot-spots, please see  \citep{SSD} for a similar concept used for image defect detection.

    To be more concrete, we assume that global trend mean and local hot-spots can be represented as $\boldsymbol{\mu}=\mathbf{B}_{m}\boldsymbol{\theta}_{m}$ and $\mathbf{h}=\mathbf{B}_{h}\boldsymbol{\theta}_{h}$, where  $\mathbf{B}_{m}$ and $\mathbf{B}_{h}$ are two bases that will be discussed below.
    The vectors $\boldsymbol{\theta}_{m}$ and $\boldsymbol{\theta}_{h}$ are the model coefficients vector of length $n_{1}n_{2}T$ and needed to be estimated, and we will discuss the estimation method later.
    Here the subscript of \textit{m} and \textit{h} are abbreviations for the mean and hot-spots.
    Since the first parameter $\boldsymbol{\theta}_{m}$ is to estimate the global trend mean, and we refer it as \textit{global mean parameter}.
    And the second parameter $\boldsymbol{\theta}_{h}$ is to estimate the local hot-spots, and we call it as \textit{local hot-spots parameter}.

    It is useful to discuss how to choose the bases $\mathbf{B}_{m}$ and $\mathbf{B}_{h},$ so as to characterize  the complex ``within'' and ``between'' correlation or interaction structures. For the ``within" correlation structures, we propose to use pre-specified bases, $\mathbf{B}_{m,s}$ and $\mathbf{B}_{h,s}$, for within-state correlation in the global trend mean and hot-spot, where the subscript of \textit{s} is an abbreviation for states.
    Similarly, $\mathbf{B}_{m,r}$ and $\mathbf{B}_{h,r}$ are the pre-specifed bases for within-correlation of the same type of crime rates, whereas $\mathbf{B}_{m,t}$ and $\mathbf{B}_{h,t}$ are the bases for within-year correlation over time.
    As for the ``between'' interaction, we use a tensor product to describe this interaction, i.e, $\mathbf{B}_{m}=\mathbf{B}_{m,s}\otimes\mathbf{B}_{m,r}\otimes\mathbf{B}_{m,t}$ and $\mathbf{B}_{h}=\mathbf{B}_{h,s}\otimes\mathbf{B}_{h,r}\otimes\mathbf{B}_{h,t}$.
    This Kronecker product has been proved to have better computational efficiency in the tensor response data \cite[see][]{kolda2009tensor}.
    With the well-structured ``within'' and ``between'' interaction, our proposed model can be written as:
    \begin{equation}
    \label{equ: model}
       \mathbf{y}
       =
       (\mathbf{B}_{m,s} \otimes \mathbf{B}_{m,r} \otimes \mathbf{B}_{m,t}) \boldsymbol{\theta}_{m}
       +
       (\mathbf{B}_{h,s} \otimes \mathbf{B}_{h,r} \otimes \mathbf{B}_{h,t}) \boldsymbol{\theta}_{h}
       +
       \mathbf{e},
    \end{equation}
    where $\mathbf{e}{\sim}N(0,\sigma^{2}\mathbf{I})$ is the random noise.
    Mathematically speaking, both $\mathbf{B}_{m,s}$ and $\mathbf{B}_{h,s}$ are $n_{1}\times n_{1}$ matrix, $\mathbf{B}_{m,r}$ and $\mathbf{B}_{h,r}$ are $n_{2}\times n_{2}$ matrix and $\mathbf{B}_{m,t}$ and $\mathbf{B}_{h,t}$ are $n_{T}\times n_{T}$ matrix, respectively.
    Besides, our proposed model in \eqref{equ: model} can also be rewritten into a tensor format:
    \begin{equation}
    \label{equ: meaning of theta}
       \mathcal{Y}
       = \vartheta_{m}\times_{3}\mathbf{B}_{m,t}\times_{2}\mathbf{B}_{m,r}\times_{1}\mathbf{B}_{m,s}+\vartheta_{h}\times_{3}\mathbf{B}_{h,t}\times_{2}\mathbf{B}_{h,r}\times_{1}\mathbf{B}_{h,s}+\mathbf{e},
    \end{equation}
    where $\vartheta_{m}$ and $\vartheta_{h}$ is the tensor format of $\boldsymbol{\theta}_{m}$ and $\boldsymbol{\theta}_{h}$ with dimensional  $n_{1}\times n_{2}\times n_{T}$.
    Accordingly, the $(i,j,t)^{th}$ element in $\vartheta_{m}, \vartheta_{h}$ ( or equivalently the $((k-1)n_{1}n_{2}+(i-1)n_{1}+j)^{th}$ entry of $\boldsymbol{\theta}_{h}$, $\boldsymbol{\theta}_{m}$) can estimate the global mean and hot-spots in the  $i^{th}$ state and $j^{th}$ crime rate in $k^{th}$ year respectively.
    The tensor representation in equation \eqref{equ: meaning of theta} allows us to develop computationally efficient methods for estimation and prediction.

    After developing the models above in equation \eqref{equ: model}, we now discuss how to  estimate the parameters $\boldsymbol{\theta}$s (the global mean parameter $\boldsymbol{\theta}_{m}$ and the local hot-spots parameter $\boldsymbol{\theta}_{h}$) in our model from the data via the penalized likelihood function.
    We propose to add two penalties in our estimation.
    First, because hot-spots rarely occur, we assume that the local hot-spots parameter $\boldsymbol{\theta}_{h}$ is sparse and the majority of entries in the hot-spot coefficient $\boldsymbol{\theta}_{h}$ are zeros.
    Thus, we propose to add the penalty term
    $
    R_{1}(\boldsymbol{\theta}_{h})=\lambda\Vert\boldsymbol{\theta}_{h}\Vert_{1}
    $
    to encourage the sparsity property of $\boldsymbol{\theta}_{h}$.
    Second, we assume the hot-spots are temporal consistent, as the unusual phenomenon of last year is likely to affect the performance of hot-spots in the current year.
    Thus, we add the second penalty
    $
    R_{2}(\boldsymbol{\theta}_{h}) = \lambda_{2}\sum_{t=2}^{T}\Vert\boldsymbol{\theta}_{h,t}-\boldsymbol{\theta}_{h,t-1}\Vert_{1}
    $
    to ensure the temporal continuity of the hot-spot, where $\boldsymbol{\theta}_{h,t}$ is a sub-vector of length $n_{1}n_{2}$ starting from the $((t-1)n_{1}n_{2}+1)^{th}$ element to the $(tn_{1}n_{2})^{th}$ element in $\boldsymbol{\theta}_{h}$, which represents the hot-spot parameter for the $t^{th}$ year.
    By combining both two penalties, we propose to estimate the parameters ($\boldsymbol{\theta}_{m}, \boldsymbol{\theta}_{h}$) via the following optimization problem:
    \begin{align}
    \arg\min_{\boldsymbol{\theta}_{m},\boldsymbol{\theta}_{h}}\Vert\boldsymbol{e}\Vert^{2}+\lambda_{1}\Vert\boldsymbol{\theta}_{h}\Vert_{1}+\lambda_{2}\sum_{t=2}^{T}\Vert\boldsymbol{\theta}_{h,t}-\boldsymbol{\theta}_{h,t-1}\Vert_{1}\nonumber \\
    s.t.\;\;\boldsymbol{y}=(\mathbf{B}_{m,s}\otimes\mathbf{B}_{m,r}\otimes\mathbf{B}_{m,t})\boldsymbol{\theta}_{m}+(\mathbf{B}_{h,s}\otimes\mathbf{B}_{h,r}\otimes\mathbf{B}_{h})\boldsymbol{\theta}_{h}+\mathbf{e},
    \label{equ: eatimation}
    \end{align}
    where $\boldsymbol{\theta}_{m}=\mathrm{vec}(\boldsymbol{\theta}_{m,1},\cdots\boldsymbol{\theta}_{m,T})$ and $\boldsymbol{\theta}_{h}=\mathrm{vec}(\boldsymbol{\theta}_{h,1},\cdots\boldsymbol{\theta}_{h,T})$.
    Combining these two penalties in equation \eqref{equ: eatimation}, we get
    $
    R(\boldsymbol{\theta}_{h})
    =
    R_{1}(\boldsymbol{\theta}_{h}) + R_{2}(\boldsymbol{\theta}_{h})
    =
    \lambda_{1}\Vert\boldsymbol{\theta}_{h}\Vert_{1}
    +
    \lambda_{2}\sum_{t=2}^{T}\Vert\boldsymbol{\theta}_{h,t}-\boldsymbol{\theta}_{h,t-1}\Vert_{1}
    $,
    which is a fussed LASSO penalty \cite[see][]{tibshirani2005sparsity} controlling both the sparsity and temporal consistency of the hot-spots.
    We will discuss how to efficiently optimize equation \eqref{equ: eatimation} for tensors in Section \ref{sec: Optimization Algorithm}.

\subsection{Optimization Algorithm for Estimation }
\label{sec: Optimization Algorithm}

    In this section, we  develop an efficient  optimization algorithm for solving the optimization problem in equation \eqref{equ: eatimation}.
    For notation convenience, we slightly adjust the notation above.
    Because $\boldsymbol{\theta}_{m},\boldsymbol{\theta}_{h}$ in equation \eqref{equ: eatimation} is solved under penalty
    $
    \lambda_{1}R_{1}(\boldsymbol{\theta}_{h})+\lambda_{2}R_{2}(\boldsymbol{\theta}_{h})
    $,
    we re-denote $\boldsymbol{\theta}_{m}$, $\boldsymbol{\theta}_{h}$ as
    $
    \boldsymbol{\theta}_{m,\lambda_{1},\lambda_{2}},\boldsymbol{\theta}_{h,\lambda_{1},\lambda_{2}}
    $
    to emphasis the penalty parameter $\lambda_{1}$ and $\lambda_{2}$.
    Accordingly, $\boldsymbol{\theta}_{h,0,\lambda_{2}}$ refers to the estimator only under the second penalty
    $
    \lambda_{2}R_{2}(\boldsymbol{\theta}_{h})
    $, i.e,
    \begin{equation}
        \boldsymbol{\theta}_{h,0,\lambda_{2}}
        =
        \arg\min_{\boldsymbol{\theta}_{m},\boldsymbol{\theta}_{h}}\{\Vert\mathbf{e}\Vert_{2}^{2} + \lambda R_{2}(\boldsymbol{\theta}_{h})\}.\label{equ: one penalty}
    \end{equation}
    The main idea of our proposed estimation algorithm can be summarized as follows.
    First we reduce the number of unknown vectors in our model, i.e, find a closed-form correlation between $\boldsymbol{\theta}_{m,\lambda_{1},\lambda_{2}}$ given $\boldsymbol{\theta}_{h,\lambda_{1},\lambda_{2}}$.
    After reducing the number of parameters, we focus on estimating $\boldsymbol{\theta}_{h,0,\lambda_{2}}$, where FISTA \cite[see][]{FISTA} is the main tool in this stage.
    We use notation $\widehat{\boldsymbol{\theta}}_{h,0,\lambda_{2}}$ to describe the corresponding estimator of $\boldsymbol{\theta}_{h,0,\lambda_{2}}$.
    Finally, the estimated
    $\boldsymbol{\theta}_{h,\lambda_{1},\lambda_{2}}$,
    described by
    $\widehat{\boldsymbol{\theta}}_{h,\lambda_{1},\lambda_{2}}$,
    is solved by the closed-form of $\widehat{\boldsymbol{\theta}}_{h,\lambda_{1},\lambda_{2}}$ given $\widehat{\boldsymbol{\theta}}_{h,0,\lambda_{2}}$.

    Following the main idea above, we first reduce the number of unknown vectors.
    Although there are two sets of parameters, namely
    $\boldsymbol{\theta}_{m,\lambda_{1},\lambda_{2}}$ and
    $\boldsymbol{\theta}_{h,,\lambda_{1},\lambda_{2}}$
    in the model, we note that given
    $\boldsymbol{\theta}_{h,\lambda_{1},\lambda_{2}}$,
    the parameter
    $\boldsymbol{\theta}_{m,\lambda_{1},\lambda_{2}}$
    is involved in the standard least squared estimation and thus can be solved in the closed-form solution, see equation \eqref{equ:theta and theta_a} in the proposition below.

    \begin{proposition}
    In the optimization problem shown in equation \eqref{equ: eatimation}, when given $\boldsymbol{\theta}_{h,\lambda_{1},\lambda_{2}}$, the closed-form solution of $\boldsymbol{\theta}_{m,\lambda_{1},\lambda_{2}}$  is given by:
    \begin{equation}
    \label{equ:theta and theta_a}
        \boldsymbol{\theta}_{m,\lambda_{1},\lambda_{2}}
        =
        (\mathbf{B}_{m}'\mathbf{B}_{m})^{-1}
        \left(\mathbf{B}_{m}'\mathbf{y} - \mathbf{B}_{m}'\mathbf{B}_{h} \boldsymbol{\theta}_{h,\lambda_{1},\lambda_{2}}\right).
    \end{equation}
    \end{proposition}

    To estimate the parameter $\boldsymbol{\theta}_{h,\lambda_{1},\lambda_{2}}$,  we plug equation \eqref{equ:theta and theta_a} into equation \eqref{equ: eatimation}.
    Then, the optimization problem for estimating $\boldsymbol{\theta}_{h,\lambda_{1},\lambda_{2}}$ in equation \eqref{equ: eatimation} becomes
    \begin{align}
    \label{equ: estimation2}
    \begin{split}
    \arg\min_{\boldsymbol{\theta}_{h,\lambda_{1},\lambda_{2}}}\Vert\mathbf{y}^{*}-\mathbf{X}\boldsymbol{\theta}_{h,\lambda_{1},\lambda_{2}}\Vert_{2}^{2}
    +
    \lambda_{1}\Vert\boldsymbol{\theta}_{h,\lambda_{1},\lambda_{2}}\Vert_{1}+\lambda_{2}\sum_{t=2}^{T}\Vert\boldsymbol{\theta}_{h,t,\lambda_{1},\lambda_{2}}
    -
    \boldsymbol{\theta}_{h,t-1,\lambda_{1},\lambda_{2}}\Vert_{1},
    \end{split}
    \end{align}
    where
    $
        \mathbf{y}^{*} = \left[\mathbf{I}-\mathbf{H}_{m}\right]\mathbf{y}
    $ ,
    $
    \mathbf X = \left[\mathbf{I}-\mathbf{H}_{m}\right]\mathbf{B}_{h}
    $
    and
    $
    \mathbf{H}_{m} = \mathbf{B}_{m}(\mathbf{B}_{m}'\mathbf{B}_{m})^{-1}\mathbf{B}_{m}'
    $
    is the projection matrix.
    Because solving the inverse of a matrix $\mathbf{B}_{m}'\mathbf{B}_{m}$ is computational expensive, therefore, in order to simply the calculation, we developed the tensor format of $\mathbf{y}^{*}$, which can be rewritten as
    $
    \mathbf{y}^{*} = \mathbf{y} - vec\left(\mathcal{Y}\times_{1}\mathbf{H}_{m,s}\times_{2}\mathbf{H}_{m,r}\times_{3}\mathbf{H}_{m,t} \right)
    $,
    where $\times_{k} (k = 1, 2, 3)$ is the mode-$k$ product in Section \ref{sec: data} and $vec(\cdot)$ is an operator to transform variable to vectors.
    And
    $\mathbf{H}_{m,s}=\mathbf{B}_{m,s}(\mathbf{B}_{m,s}'\mathbf{B}_{m,s})^{-1}\mathbf{B}_{m,s}'$, $\mathbf{H}_{m,r}=\mathbf{B}_{m,r}(\mathbf{B}_{m,r}'\mathbf{B}_{m,r})^{-1}\mathbf{B}_{m,r}'$, and $\mathbf{H}_{m,t}=\mathbf{B}_{m,t}(\mathbf{B}_{m,t}'\mathbf{B}_{m,t})^{-1}\mathbf{B}_{m,t}'$.
    The details of the proof are shown in Appendix \ref{proof: prop 1}.

To develop an efficient optimization algorithm to solve the global optimum of equation \eqref{equ: eatimation}, we first solve
    $
    \boldsymbol{\theta}_{h,0,\lambda_{2}}
    $,
    i.e., with the sparsity penalty parameter $\lambda_1= 0$, which is used to solve
    $
    \boldsymbol{\theta}_{h,\lambda_1,\lambda_2}
    $
    for general $\lambda_1$ later.
    The estimation  of $\boldsymbol{\theta}_{h,0,\lambda_{2}}$ is to optimize
    \begin{equation}
        \arg\min_{\boldsymbol{\theta}_{h,0,\lambda_{2}}}
        \Vert\mathbf{y}^{*}-\mathbf{X}\boldsymbol{\theta}_{h,0,\lambda_{2}}\Vert_{2}^{2}
        +
        \lambda_{2}\Vert\boldsymbol{D\theta}_{h,0,\lambda_{2}}\Vert_{1},
        \label{equ: estimate theta0lambda2}
    \end{equation}
    where matrix $\mathbf D$ is of dimension $n_{1}n_{2}(n_{3}-1)\times n_{1}n_{2}n_{3}$, whose $(i,i)^{th}$, $(i,i+n_{1}n_{2}-1)^{th}$ elements are of value $-1$ and $1,$ respectively.

    Obviously, the optimization problem for equation \eqref{equ: estimate theta0lambda2} is a generalized LASSO problem, which can be transformed into a regular LASSO problem \cite[see][]{generalizedLasso}.
    The details of the transformation procedure are shown in  Proposition \ref{prop: generalized lasso to lasso}.

    \begin{proposition}
    \label{prop: generalized lasso to lasso}
    Assume that
    \begin{enumerate}
    \item
        matrix $\mathbf A$ is of dimension $n_{1}n_{2} \times n_{1}n_{2}n_{3}$ and constructed through combining $n_{1}n_{2} \times n_{1}n_{2}$ identity matrix rowly $n_{3}-1$ times,
    \item
        matrix $\widetilde{\mathbf D}$ of dimension $n_{1}n_{2}n_3\times n_{1}n_{2}n_{3}$ and is defined as
        $
        \widetilde{\mathbf D}
        =
        \begin{bmatrix}
        \mathbf D \\
        \mathbf A
        \end{bmatrix}
        $,
    \item
        matrix $\mathbf{X}_{1}$ is the first $n_{1}n_{2}(T-1)$ rows of matrix $\mathbf X\widetilde{\mathbf D}^{-1}$, matrix $\mathbf X_{2}$ is the remaining part of matrix $\mathbf X\widetilde{\mathbf D}^{-1}$,
    \item
        the projection onto the column space of $\mathbf X_2$ is noted as
        $
        \mathbf P = \mathbf X_{2}(\mathbf X_{2}'\mathbf X_{2})^{-1}\mathbf X_{2}'
        $,
    \item
        the first $n_1n_2(n_3-1)$ entries of $\boldsymbol{\beta}$ ($\boldsymbol{\beta}=\widetilde{\mathbf D}\boldsymbol{\theta}_{h,0,\lambda_2}$ is noted as $\boldsymbol{\beta}_1$, and the remaining are noted as $\boldsymbol{\beta}_2$.
    \end{enumerate}
    The generalized LASSO problem in equation \eqref{equ: estimate theta0lambda2} can be solved by $\widehat{\boldsymbol\theta}_{h,0,\lambda_{2}}=\widetilde{\mathbf D}^{-1}\hat{\boldsymbol\beta},$ where $\widehat{\boldsymbol\beta}=(\widehat{\boldsymbol\beta}_1,\widehat{\boldsymbol\beta}_2)$ and
	\begin{eqnarray}
	\label{equ: generalized lasso to lasso}
        \widehat{\boldsymbol\beta}_{1}
        &=&
        \arg\min_{\boldsymbol{\beta}_1}
        \Vert\mathbf{(\mathbf I-\mathbf P)y}^{*}-(\mathbf I-\mathbf P)\mathbf{X_{1}}\boldsymbol\beta_{1}\Vert_{2}^{2}
        +
        \lambda_{2}\Vert\boldsymbol \beta_{1}\Vert_{1}\\
        \widehat{\boldsymbol\beta}_{2}&=&(\mathbf X_{2}'\mathbf X_{2})^{-1}\mathbf X_{2}'(\mathbf y^*-\mathbf X_{1}\widehat{\boldsymbol\beta}_{1}).\nonumber
	\end{eqnarray}
    \end{proposition}
    Proposition \ref{prop: generalized lasso to lasso} allows us to focus on efficiently solving the LASSO-type optimization problem in \eqref{equ: generalized lasso to lasso}.
    To do so, we propose to use the  FISTA algorithm in \cite{FISTA} due to its fast convergence rate.
    Indeed, \cite{FISTA} showed that the convergence rate of the FISTA algorithm is of order $O(1/k^2)$ where $k$ indicates the iterations.
    FISTA is a very efficient and standardized algorithm in optimization research to solve the optimization problem involving $\ell_1$ penalty.
    The proof of Proposition \ref{prop: generalized lasso to lasso} can be found in Appendix \ref{proof: prop 2}.

    Finally, with the solved $\widehat{\boldsymbol{\theta}}_{h,0,\lambda_{2}}$ for $\lambda_1 = 0 $, we can easily compute
    $\widehat{\boldsymbol{\theta}}_{h,\lambda_1,\lambda_{2}}$ for general $\lambda_1 > 0$ due to the their closed form relations in the following proposition:

    \begin{proposition}
    \label{prop: theta_0_lambda2 and theta_lambda1_lambda2}
    The closed form relationship between
    $\widehat{\boldsymbol{\theta}}_{h,\lambda_{1},\lambda_{2}}$ and
    $\widehat{\boldsymbol{\theta}}_{h,0,\lambda_{2}}$ is
    \begin{align}
    \label{equ: corre between lambda1 and lambda2}
        \widehat{\boldsymbol{\theta}}_{h,\lambda_{1},\lambda_{2}}
        =
        sign(\widehat{\boldsymbol{\theta}}_{h,0,\lambda_{2}})
        \odot
        \max\{|\widehat{\boldsymbol{\theta}}_{h,0,\lambda_{2}}| - \lambda_{1},0 \}.
    \end{align}
    where $\odot$ is an element-wise product operator.
    \end{proposition}
    The proof of Proposition \ref{prop: theta_0_lambda2 and theta_lambda1_lambda2} could be found in the proof of Theorem 1 of \cite{liu2010efficient}.
    In summary, the details of our proposed optimization algorithm are shown in Algorithm \ref{alg: estimate beta1} below.

    \begin{algorithm}[H]
    \label{alg: estimate beta1}
    \caption{ Estimation of $\boldsymbol{\theta}_{h,\lambda_{1},\lambda_{2}}$}
    \LinesNumbered	
	\KwIn{$\mathbf{y}$,
          $\mathbf{B}_{m,s},\mathbf{B}_{m,r},\mathbf{B}_{m,t}$,
          $\mathbf{B}_{h,s},\mathbf{B}_{h,r},\mathbf{B}_{h,t}$,
          $\mathbf X_{1}, \mathbf X_{2}, $
          $\widetilde{\mathbf X} = (\mathbf{I} - \mathbf{P})\mathbf{X}_1$,
          $\widetilde{\mathbf y} = (\mathbf{I} - \mathbf{P})\mathbf{y}^*$,
          $\lambda_1, \lambda_2, K$}
	\KwOut{$\widehat{\boldsymbol{\theta}}_{h,\lambda_{1},\lambda_{2}}$}
	\bfseries{initialization}\;  	
	      $\boldsymbol\beta_1^{(0)}$,
          $\boldsymbol\alpha^{(1)} = \boldsymbol\beta_1^{(0)}$ ,
          $t_1=1$, $k=0$ \\
	\For{ $k= 1\cdots K$ }{
          $
          \boldsymbol\beta_1^{(k)}
          =
          S
          \left(
          \boldsymbol\alpha^{(k)}
          -
          \frac{1}{nL}
          \left(
          \widetilde{\mathbf X}'\widetilde{\mathbf X} \boldsymbol{\alpha}^{(k)}
          +
          \widetilde{\mathbf X}'\widetilde{\mathbf y}
          \right),\lambda_2/L
          \right)
          $ \\
          \tcc{$S(\cdot)$ is the soft-thresholding function.}
          $
          t_{k+1} = \frac{ 1 + \sqrt{1 + 4 t_k^2}}{2}
          $  \\
          $
          \boldsymbol\alpha^{(k+1)}
          =
          \boldsymbol\beta^{(k)}
          +
          \frac{t_k-1}{t_{k+1}}
          \left(
          \boldsymbol\beta^{(k)}-\boldsymbol\beta^{(k-1)}
          \right)
          $   \\
          $ k = k + 1 $
	}
    $
    \widehat{\boldsymbol\beta}_{2}
    =
    (\mathbf X_{2}'\mathbf X_{2})^{-1}\mathbf X_{2}'(\mathbf y-\mathbf X_{1}\boldsymbol\beta_{1}^{(K)})
    $\\
    $
    \widehat{\boldsymbol\theta}_{h,0,\lambda_{2}}
    =
    \widetilde{\mathbf D}^{-1}(\boldsymbol\beta_{1}^{(K)},\widehat{\boldsymbol\beta}_{2})'
    $ \\
    $
    \widehat{\boldsymbol{\theta}}_{h,\lambda_{1},\lambda_{2}}
    =
    sign(\widehat{\boldsymbol{\theta}}_{h,0,\lambda_{2}})
    \odot
    \max\{|\widehat{\boldsymbol{\theta}}_{h,0,\lambda_{2}}|-\lambda_{1},0\}
    $\\
\end{algorithm}

\subsection{Selection of Bases in Practice}
\label{sec: choice of basis}

    This section discusses how to choose the proper bases $\mathbf{B}_{m,s}$, $\mathbf{B}_{m,r}$,
    $\mathbf{B}_{m,t}$,
    $\mathbf{B}_{h,s}$,
    $\mathbf{B}_{h,r}$,
    $\mathbf{B}_{h,t}$.
    Generally speaking, some reasonable choices of the bases can be:
    (1) identity matrix when one has little to no prior knowledge of the data structure;
    (2) Gaussian kernel if the data shares a very smooth background;
    (3) Other kernels, including Cosine, Silverman and etc.,  depending on the nature or characteristics of the data.

    In the crime rate data, we begin with the basis for the global trend mean.
    Figure \ref{fig: Raw Data map} shows that the global pattern is very smooth, where no distinctive  abrupt changes between neighbor states.
    To model the smooth spatial correlation of the global state pattern, we propose to apply  kernel matrix, defined as $\mathbf{B}_{m,s}$ with $\exp\{- d^2/(2c^2)\}$ for the $(i,j)^{th}$ element, where $d$ is the distance between the $i^{th}$ state and $j^{th}$ state, and $c$ is the bandwidth chosen by cross-validation.
    For the correlation among the crime rates, we set  $\mathbf{B}_{m,r}$ to be an identity matrix, since we do not have any prior knowledge of it. Next, for the bases for the hot-spots, we assume there is no prior knowledge of the hot-spots.
    Thus we set  $\mathbf{B}_{h,s}$ and $\mathbf{B}_{h,r}$ to be an identity matrix.
    Moreover, for the temporal basis in both global trend mean and hot-spots,  the identity matrix is used, while this reflects that we do not know when hot-spots will occur. Our optimization algorithm includes an autoregressive-type regularization term in the estimation procedure to guarantee that there is the temporal continuity of hot-spot.

\section{Detection and Localization of Hot-spots}
\label{sec: Temporal Detection and Spatial Location of Hot Spot}

    This section focuses on the detection and localization of the hot-spot, which includes detecting the year (when) and localizing the state (where) as well as finding the correct crime type (which) of the hot-spot.
    In our case study, we focus on the upward shift of crime rates, since the increasing crime rates are generally more harmful to the societies and communities.
    Of course, one can also detect the downward shift with a slight modification of our proposed algorithms by multiplying $-1$ to the raw data.

    For the ease of presentation, we first discuss the detection of the hot-spot, i.e., detect when a hot-spot occurs in Section \ref{sec: Temporal Detection}.
    Then, in Section \ref{sec: spatial location}, we consider the localization of the hot-spot, i.e., determine which states and which crime types are involved for the detected hot-spot.

\subsection{Detect When the Hot Spot Occurs?}
\label{sec: Temporal Detection}

    To determine when the hot-spot occurs, we consider the following hypothesis test and set up the control chart for the hot-spot detection in equation \eqref{eq:chaneg_hypothesis_testing}.
    \begin{equation}
         H_{0}: \; \mathbf{r}_{t} = 0 \;\;\; v.s. \;\;\; H_{1}:\; \mathbf{r}_{t} = \delta\widehat{\mathbf{h}}_{t} \;\;\;(\delta>0),
         \label{eq:chaneg_hypothesis_testing}
    \end{equation}
    where $\mathbf{r}_{t}$ is the expected residuals after removing the mean.
    The essence of this test is that, we want to detect whether $\mathbf{r}_{t}$ has a mean shift in the direction of $\widehat{\mathbf{h}}_{t}$ (estimated in Section \ref{sec: Optimization Algorithm}).
    Please note that $\widehat{\mathbf{h}}_{t}$ is the subvector of $\widehat{\mathbf{h}}$, starting from the $((t-1)n_1 n_2 + 1)^{th}$ element to the $(n_1 n_2 t)^{th}$ element.
    And $\widehat{\mathbf{h}}$ is derived by $\widehat{\mathbf{h}} = \mathbf{B}_h \boldsymbol{\theta}_h$.

    To test this hypotheses, the likelihood ratio test is applied to the residual $\mathbf{r}_{t}$ at each time $t$, i.e., $\mathbf{r}_{t}=\mathbf{y}_{t}-\boldsymbol{\mu}_{t}$, where it assumes that the residuals $\mathbf{r}_{t}$ is independent after removing the mean and its distribution before and after the hot-spot remains the same.
    Accordingly, the test statistics monitoring upward shift is designed as
    $
    P_{t}^{+}=\widehat{\mathbf{h}}_{t}'^{+}\mathbf{r}_{t} \bigg/ \sqrt{\widehat{\mathbf{h}}_{t}'^{+}\widehat{\mathbf{h}}_{t}^{+}}
    $
    \cite[see][]{hawkins1993regression},
    where $\widehat{\mathbf{h}}_{t}^{+}$ only takes the positive part of $\widehat{\mathbf{h}}_{t}$ with other entries as zero. Here we put a
    superscript ``+'' to emphasis that it aims for upward shift.

    Unfortunately, different choices of the penalty parameters $\lambda_{1},\lambda_{2}$ gives different test statistics $P_{t}^{+}$.
    In order to select the one with the most power, we propose to calculate a series of $P_{t}^{+}$ under different combination of $(\lambda_{1},\lambda_{2})$ from the set $\Gamma=\{(\lambda_{1}^{(1)},\lambda_{2}^{(1)})\cdots(\lambda_{1}^{(n_{\lambda})},\lambda_{2}^{(n_{\lambda})})\}$.
    For better illustration, we denote the test statistics under the penalty parameters $(\lambda_{1},\lambda_{2})$ as $P_{t}^{+}(\lambda_{1},\lambda_{2})$.
    The test statistics \cite[see][]{zou2009multivariate} with the most power to detect the change, noted as $\tilde{P}_{t}^{+}$, can be computed by
    \begin{equation}
        \widetilde{P}_{t}^{+}
        =
        \max_{(\lambda_{1},\lambda_{2})\in\Gamma}
        \frac{P_{t}^{+}(\lambda_{1},\lambda_{2})-E(P_{t}^{+}
        (\lambda_{1},\lambda_{2}))}{\sqrt{Var(P_{t}^{+}(\lambda_{1},\lambda_{2}))}},
    \label{equ: most power}
    \end{equation}
    where $E(P_{t}^{+}(\lambda_{1},\lambda_{2}))$, $Var(P_{t}^{+}(\lambda_{1},\lambda_{2}))$ respectively are the mean and variance of $P_{t}(\lambda_{1},\lambda_{2})$ under $H_{0}$ (e.g., for phase-I in-control samples).

    Note that the penalty parameters $(\lambda_{1},\lambda_{2})$ detect maximization in equation \eqref{equ: most power} is generally different under different times $t$.
    To emphasize such dependence of time $t$, we denote the parameter pair that attains the maximization in equation \eqref{equ: most power} at time $t$ as $(\lambda_{1,t}^{*},\lambda_{2,t}^{*})$,  i.e,
    \begin{equation}
    \label{eq:lambda12}
        (\lambda_{1,t}^{*},\lambda_{2,t}^{*})
        =
        \arg\max_{(\lambda_{1},\lambda_{2})\in\Gamma}
        \frac{P_{t}^{+}(\lambda_{1},\lambda_{2})-E(P_{t}^{+}(\lambda_{1},\lambda_{2}))}{\sqrt{Var(P_{t}^{+}(\lambda_{1},\lambda_{2}))}}.
    \end{equation}
    Thus, the series of the test statistics for the hot-spot at time $t$ is $\widetilde{P}_{t}^{+}(\lambda_{1,t}^{*},\lambda_{2,t}^{*})$ where $t=1\cdots T$.

    With the test statistic available, we design a control chart based on the CUSUM procedure for the following reasons:
    1) we are interested in detecting the change with the temporal continuity, therefore,  alignment with the objective of CUSUM.
    2) In the view of social stability, we want to keep the crime rates at a target value without sudden changes, which makes the CUSUM chart is a naturally better fit.

    Specifically, in the CUSUM procedure, we compute the CUSUM statistics recursively by
    $$
       W_{t}^{+}=\max\{0,W_{t-1}^{+}+\widetilde{P}_{t}^{+}(\lambda_{1,t}^{*},\lambda_{2,t}^{*})-d\},
    $$
    and
    $
    W_{t=0}^{+}=0
    $, where $d$ is a constant and can be chosen according to the degree of the shift that we want to detect.
    Next, we set the control limit $L$ as the four times of the standard derivation of $\widetilde{P}_{t}^{+}(\lambda_{1,t}^{*},\lambda_{2,t}^{*})(t=1\cdots T)$.
    Finally, whenever $W_{t}^{+} > L$ at some time $t=t^{*},$ we determine that a  hot-spot occurs at time $t^{*}$.


\subsection{Localize Where and Which the Hot Spot Occur?}
\label{sec: spatial location}

    After the hot-spot $t^*$ has been detected by the CUSUM control chart in the previous subsection, the next step is to localize  where and which crime rate may account for this hot-spot.
    To do so, we propose to utilize the  vector
    $$
    \widehat{\mathbf{h}}_{\lambda_{1, t^{*}}^{*}, \lambda_{2,t^{*}}^{*}}
    =
    \mathbf{B}_{h}\widehat{\boldsymbol{\theta}}_{h, \lambda_{1,t^{*}}^{*}, \lambda_{2,t^{*}}^{*}}
    $$
    at the declared hot-spot time $t^{*}$ and the corresponding  parameter
    $\lambda_{1,t^{*}}^{*},\lambda_{2,t^{*}}^{*}$
    in equation \eqref{eq:lambda12}.
    For the numerical computation purpose, it  is often easier to directly work with the tensor format of the hot-spot
    $
    \widehat{\mathbf{h}}_{ \lambda_{1,t^{*}}^{*},\lambda_{2,t^{*}}^{*}}
    $,
    denoted as
    $
    \widehat{\mathcal{H}}_{\lambda_{1,t^{*}}^{*}, \lambda_{2,t^{*}}^{*}}
    $,
    which is a tenor of dimension $ n_{1} \times n_{2} \times T $.
    If the $(i,j, t^*)^{th}$ entry in $\widehat{\mathcal{H}}_{\lambda_{1,t^{*}}^{*}, \lambda_{2,t^{*}}^{*}}$ is non-zero, then we declare that there is a  hot-spot for the $j^{th}$ crime rate type in the $i^{th}$ state in $t^{*th}$ year.

\section{Simulation Study}
\label{sec: simulation}
    In this section, we conduct simulation studies to evaluate our proposed methodologies by comparison with several benchmark methods in the literature.
    The structure of this section is as follow.
    We first present the data generation mechanism for our simulations in Section \ref{sec: sim data generation}, then discuss the performance of hot-spot detection and localization in Section \ref{sec: hot-spot dectetion performance}, and finally investigate the fitness of the global trend mean in Section \ref{sec: sim_background fitness}.

\subsection{Data Generation}
\label{sec: sim data generation}
    In our simulation, at each time index $t (t=1 \cdots T)$,  we generate a vector $\mathbf y_t$ of length $n_{1} n_{2} $ by
    \begin{equation}
    \label{equ: sim data generation}
        \mathbf y_{i,t}
        =
        (\mathbf B\boldsymbol\theta_t)_i
        +
        \delta\mathbbm 1 \{t\geq \tau\} \mathbbm 1_i\{i \in S_h\}
        +
        \mathbf w_{i,t},
    \end{equation}
    where  $\mathbf y_{i,t}$ denotes the $i$-th entry in vector  $\mathbf y_t$, and $(\mathbf B\boldsymbol\theta_t)_i$ denotes the $i$-th entry in vector the $\mathbf B\boldsymbol\theta_t$.
    Besides, parameter $\delta$ denotes the change magnitude.
    Here  $\mathbbm 1(A)$ is the indicator function, which has the value 1 for all elements of $A$ and  the value 0 for all elements  not in $A$, and $\mathbf w_{i,t}$ is the $i$-th entry in the white noise vector whose entries are independent and follow  $N(0,0.1^{2})$ distribution.

    For the anomaly setup,  $\mathbbm 1\{t\geq \tau\} $  indicates that the hot-spots only occur after the hot-spot $\tau$.
    This ensures that the simulated hot-spot is temporal consistent.
    The second indicator function $\mathbbm 1_i\{i \in S_h\}$ shows that only those entries whose location index belongs set $S_h$ are assigned as local hot-spots.
    This ensures that the simulated hot-spot is sparse.
    Here we assume the change happens at  $ \tau = 20$ and the hot-spots index set
    $
    S_{h} = \{3,4,5,45,46,47,57,58,59,77,78,79,119, 120, 121, 137, 138, 139\}
    $.

    To match the dimension in the case study, we choose $n_{1}=48,n_{2}=3, T=50$.
    For the three terms on the right side of equation  \eqref{equ: sim data generation}, they serve for the global trend mean, local sparse anomaly, and white noise, respectively.

    In our simulation, the matrix $\mathbf B$ is a fixed B-spline basis with the degree of three and ten knots.
    Note that the B-spline basis is only used in the generative model in simulation to generate data, but is not used in our proposed methodologies.

    Moreover, the vector $\boldsymbol\theta_t$ is  generated by a normal distribution, and we consider the following two scenarios:
    \begin{itemize}
         \item
             Scenario 1:
             The global trend mean is stationary, in which  $ \boldsymbol \theta_t$ is generated by the normal distribution with mean $ 1 $ and standard deviation $ 0.1 $.
         \item
             Scenario 2:
             The  global trend mean is decreasing over time, in which  $\boldsymbol\theta_t$ is generated by normal distribution with mean $0.95^{t-1}$ and standard deviation $0.1$.
    \end{itemize}
    Moreover, in each of these two scenarios, we further consider two subcases, depending on the value of change magnitude $\delta$ in  (\ref{equ: sim data generation}): one is $\delta = 0.1$ (small shift) and the other is $\delta=0.5$ (large shift).

\subsection{Hot-spot Detection Performance}
\label{sec: hot-spot dectetion performance}
    In this section, we compare the performance of our proposed method (denoted as `SSR-tensor') for the detection of hot-spots with some benchmark methods.
    Specifically, we compare our proposed method with
    Hotelling $T^{2}$ control chart \cite{T2} (denoted  as `T2'),
    LASSO-based control chart proposed by \cite{zou2009multivariate} (denoted as `ZQ LASSO'),
    PCA-based control chart \cite{PCA}  and
    SSD proposed by \cite{SSD}(denoted as `SSD').

    For the basis choices of  our proposed method, to model the spatial structure of the global trend mean, we choose $B_{m,1}$ as the kernel matrix to describe the smoothness of the background, whose $(i,j)$ entry is of value $\exp\{-d^2/(2c^2)\}$ where $d$ is the distance between the $i$-th state and $j$-th state and $c$ is the bandwidth chosen by cross-validation.
    In addition, we choose identity matrices for the temporal basis and crime type basis since we do not have any prior information.
    Moreover, we use the identity matrix for the spatial and temporal basis of the hot-spots.
    For SSD in \cite[see][]{SSD}, we will use the same spatial and temporal basis in order to have a fair comparison.

    For evaluation, we will compute the following four criteria:
    (i) precision, defined as the proportion of detected hot-spots that are true hot-spots;
    (ii) recall, defined as the proportion of the hot-spots that are correctly identified;
    (iii) F-measure, a single criterion that combines the precision and recall by calculating their harmonic mean; and
    (iv) the corresponding average run length ($\mbox{ARL}_1$), a measure on the average detection delay in the special scenario when the change occurs at time $t=1$.
    All  simulation results below are based on $1000$ Monte Carlo simulation replications.

    Table \ref{table: simulation_hotspot no global trend} and Table \ref{table: simulation_hotspot decreasing global trend} report the simulation results of our proposed SSR-tensor method and  four other baseline methods.
    Note that the two baseline methods, PCA and T2, cannot localize the hot-spots, and thus we do not report the corresponding values on the precision, recall, and F-measure.
    Moreover, in our simulation, if a method fails to detect any hot-spots within the entire $T=50$ years (recall that the true hot-spot occurs at time $\tau=20$), we record its detection delay as $30$.
    Thus, for those methods with a large standard deviation of $\rm{ARL}_1$, it  is likely caused by the failure of reporting hot-spots.

    From Tables \ref{table: simulation_hotspot no global trend} and \ref{table: simulation_hotspot decreasing global trend}, it is easy to see that our proposed SSR-tensor method achieves the smallest $\rm{ARL}_1$  and largest F-measure due to the ability to capture both temporal consistency and spatial sparsity of the hot-spots.
    This implies that our proposed method not only provides a more immediate alarm when the hot-spots occur but also gives a more accurate estimation of the hot-spots location when they occur.

    Meanwhile, the two baseline methods, SSD and ZQ LASSO, are worse than our proposed SSR-tensor method due to their inability to capture the temporal continuity of the hot-spots, particularly in Scenario 2 (decreasing global trend mean).
    The baseline methods, PCA and T2, perform the worst  due to their inability to detect the sparse changes.
    In particular, T2 fails to detect the hot-spots even under the large shift scenario (i.e.,  $\rm{ARL}_1$ is 30).
    The reason is that both T2 and PCA are designed based on a multivariate hypothesis test on the global mean change, which cannot consider the sparsity of the change as well as the non-stationary global mean trend.

\begin{table}[htbp]
    \centering
	\scriptsize
	\begin{tabular}{c|cccc|cccc}
		\hline
		\multicolumn{1}{c|}{ \multirow{2}{*}{methods} } & \multicolumn{4}{c|}{large shift $\delta=0.5$} & \multicolumn{4}{c}{small shift $\delta=0.1$} \\
		\cline{2-9}
		\multicolumn{1}{c|}{} & precision & recall & F measure & ARL & precision & recall & F measure & ARL \\
		\hline
		SSR-tensor
		&\bf{0.2714}   & \bf{0.9667}   & \bf{0.6190}   & \bf{1.0003}
		&\bf{0.2401}   & \bf{0.9778}   & \bf{0.6089}   & \bf{1.2130}  \\
		&(0.0171)      & (0.0286)      & (0.0188)      & (0.1026)
		&(0.0219)      & (0.0388)      & (0.0283)      & (0.3078)  \\
		SSD
		&0.2636   & 0.9840   & 0.6238   & 1.0018
		&0.2311   & 0.9300   & 0.5806   & 1.7865  \\
		&(0.0141) & (0.0292) & (0.0189) &(0.1132)
		&(0.0553) & (0.2164) & (0.1353) &(0.5693) \\
		ZQ LASSO
		&0.1351  & 0.9850   & 0.5600   & 2.4178
		&0.1325  & 0.8771   & 0.5048   & 5.9560 \\
		&(0.0180)& (0.0302) & (0.0118) & (1.0097)
		&(0.0124)& (0.1215) & (0.0586) & (1.7385)\\
		PCA
		& - & - & - & 19.8310
		& - & - & - & 24.4120\\
		& - & - & - & (9.0701)
		& - & - & - &(10.4709) \\
		T2
		& - & - & - & 30.0000
		& - & - & - & 30.0000\\
		& - & - & - & (0.0000)
		& - & - & - & (0.0000)\\
		\hline
	\end{tabular}
	\caption{Scenario 1 (stationary global trend mean): Comparison of hot-spot detection under small and large shifts
}
	\label{table: simulation_hotspot no global trend}
\end{table}

\begin{table}[htbp]
	\centering
    \scriptsize
	\begin{tabular}{c|cccc|cccc}
		\hline
		\multicolumn{1}{c|}{ \multirow{2}{*}{methods} } & \multicolumn{4}{c|}{large shift $\delta=0.5$} & \multicolumn{4}{c}{small shift $\delta=0.1$}\\
		\cline{2-9}
		\multicolumn{1}{c|}{} & precision & recall & F measure & ARL & precision & recall & F measure & ARL \\
		\hline
		SSR-tensor
		&\bf{0.3068}   & \bf{0.9999}   & \bf{0.6534}   & \bf{1.0800}
		&\bf{0.2538}   & \bf{0.9833}   & \bf{0.6186}   & \bf{9.0087} \\
		&(0.0435)      & (0.0001)      & (0.0217)      & (0.6831)
		&(0.0155)      & (0.0268)      & (0.0194)      & (5.4261)  \\
		SSD
		&0.2839   & 0.9944   &  0.6392  & 1.4300
		&0.2298   & 0.8856   &  0.5578  & 12.1600 \\
		&(0.0221) & (0.0201) & (0.0180) & (0.8072)
		&(0.0781) & (0.2981) & (0.1878) & (9.4598) \\
		ZQ LASSO
		&0.1251   & 0.9800   &0.5556    &3.4770
		&0.0457   & 0.3609   &0.2033    &20.7200 \\
		&(0.0150) & (0.1175) & (0.0662) &(4.6848)
		&(0.0607) & (0.4794) &(0.2701)  &(12.6450) \\
		PCA
		& - & - & - & 24.3220
		& - & - & - & 21.5380\\
		& - & - & - & (8.8534)
		& - & - & - & (9.2965) \\
		T2
		& - & - & - & 30.0000
		& - & - & - & 30.0000\\
		& - & - & - & (0.0000)
		& - & - & - & (0.0000)\\
		\hline
	\end{tabular}
	\caption{Scenario 2 (decreasing global trend mean): Comparison of hot-spot detection under small and large shifts
}
	\label{table: simulation_hotspot decreasing global trend}
\end{table}

\begin{figure}[htbp]
	\begin{tabular}{ccc}
		\centering
		\includegraphics[width=0.31\textwidth]
		{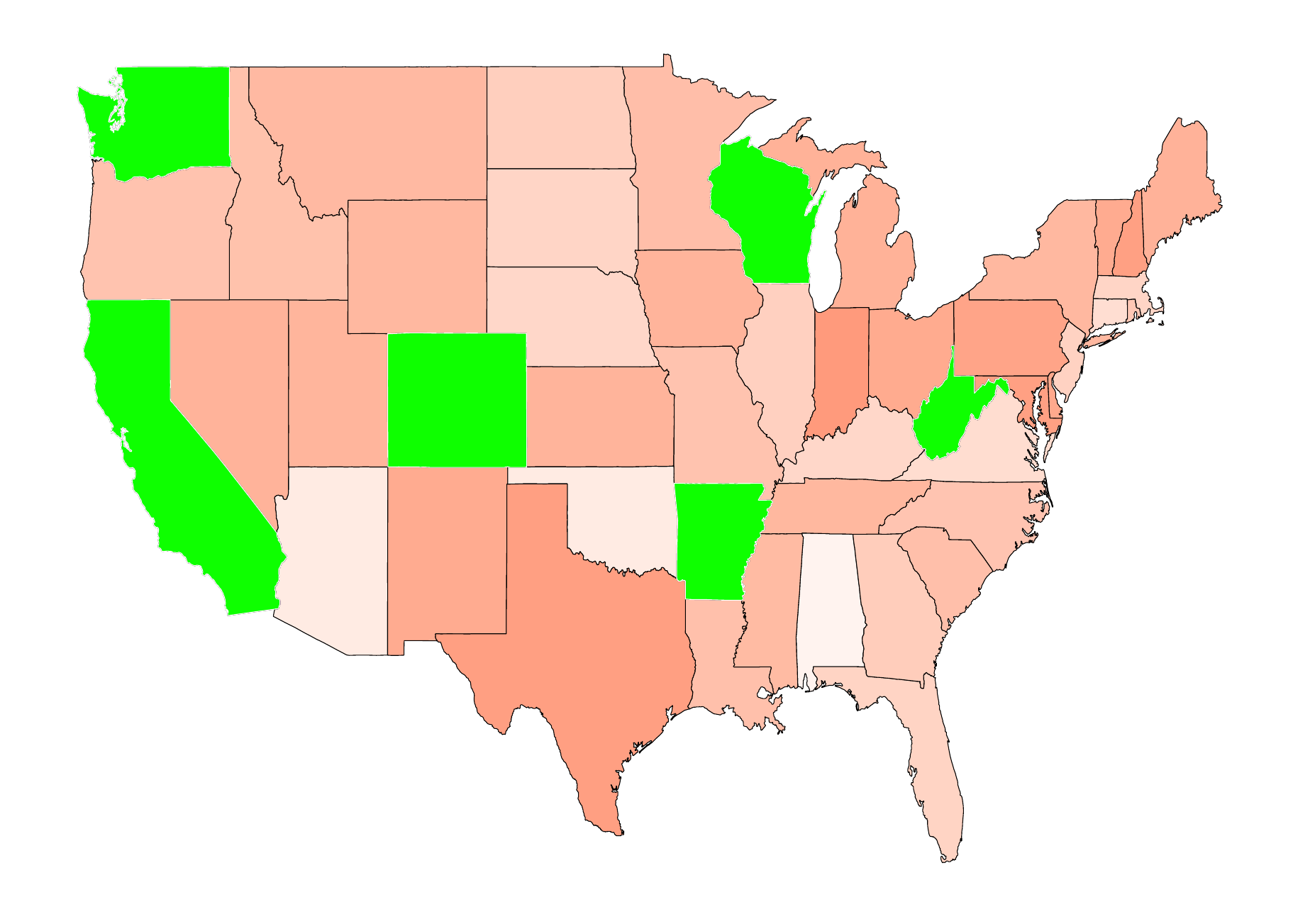}  &
		\includegraphics[width=0.31\textwidth]
		{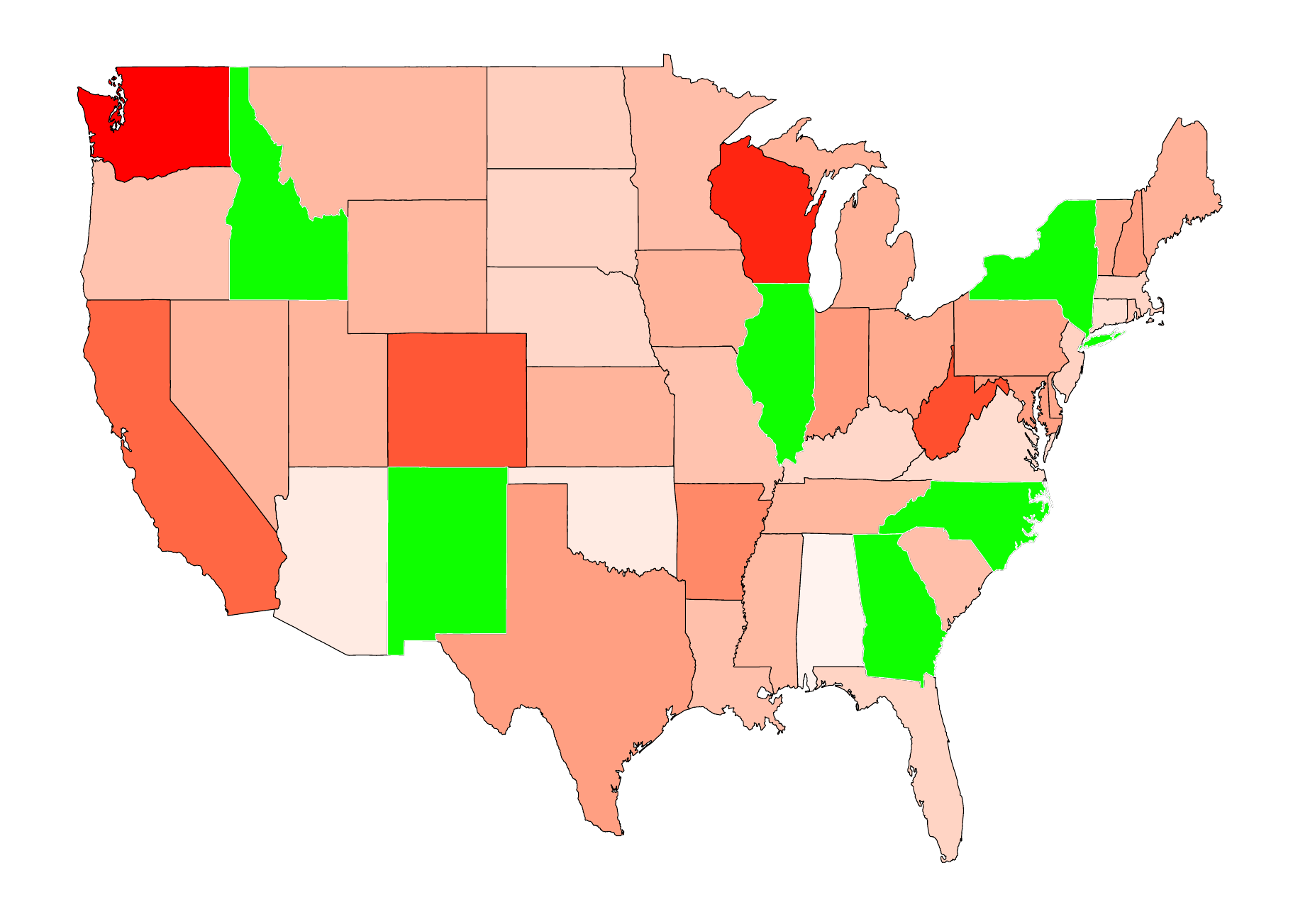}
		& \includegraphics[width=0.31\textwidth]
		{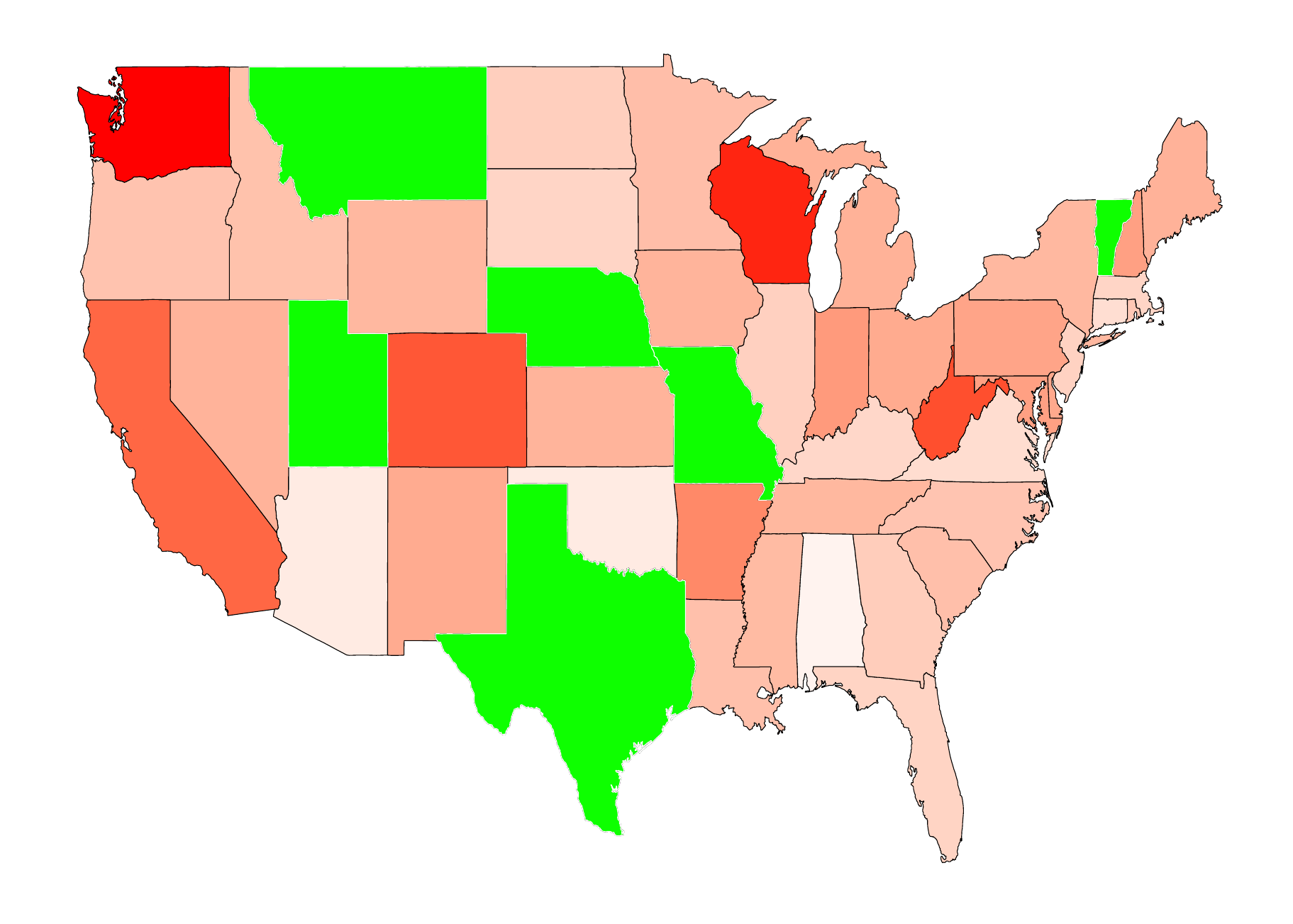} \\
		(a) ZQ LASSO in category1  &
		(b) ZQ LASSO in category2  &
		(c) ZQ LASSO in category3 \tabularnewline
		\includegraphics[width=0.31\textwidth]
		{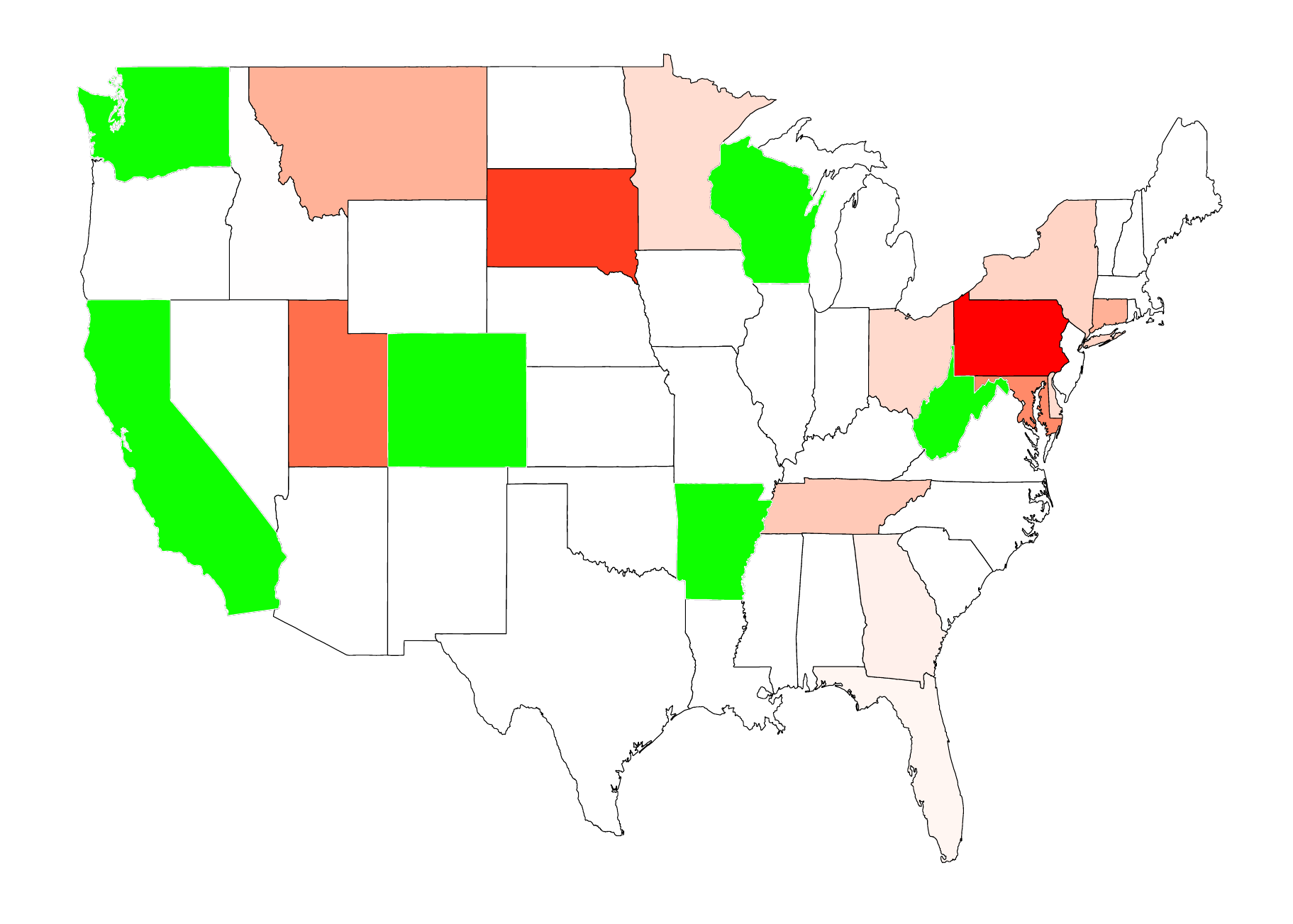}  &
		\includegraphics[width=0.31\textwidth]
		{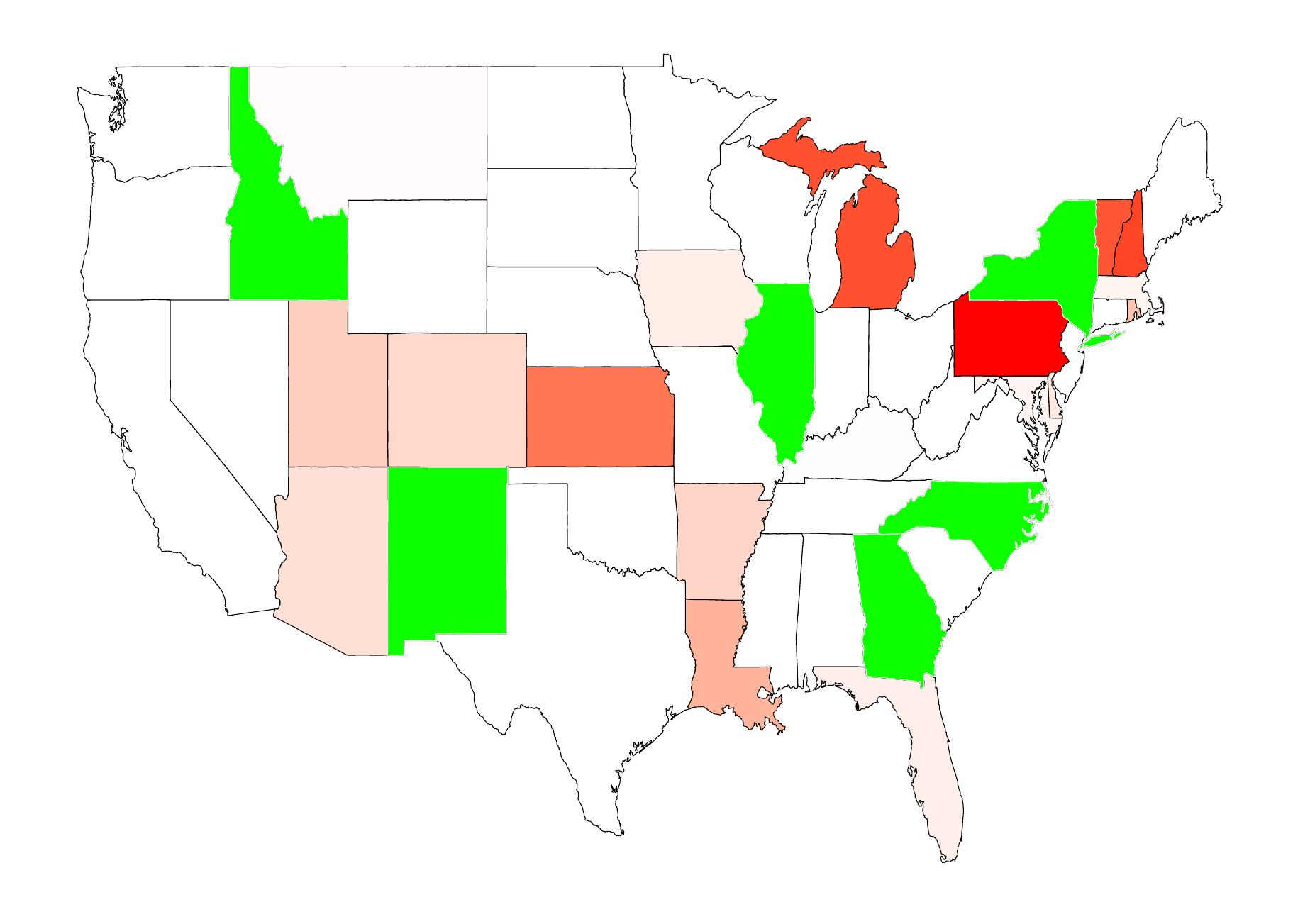}  &
		\includegraphics[width=0.31\textwidth]
		{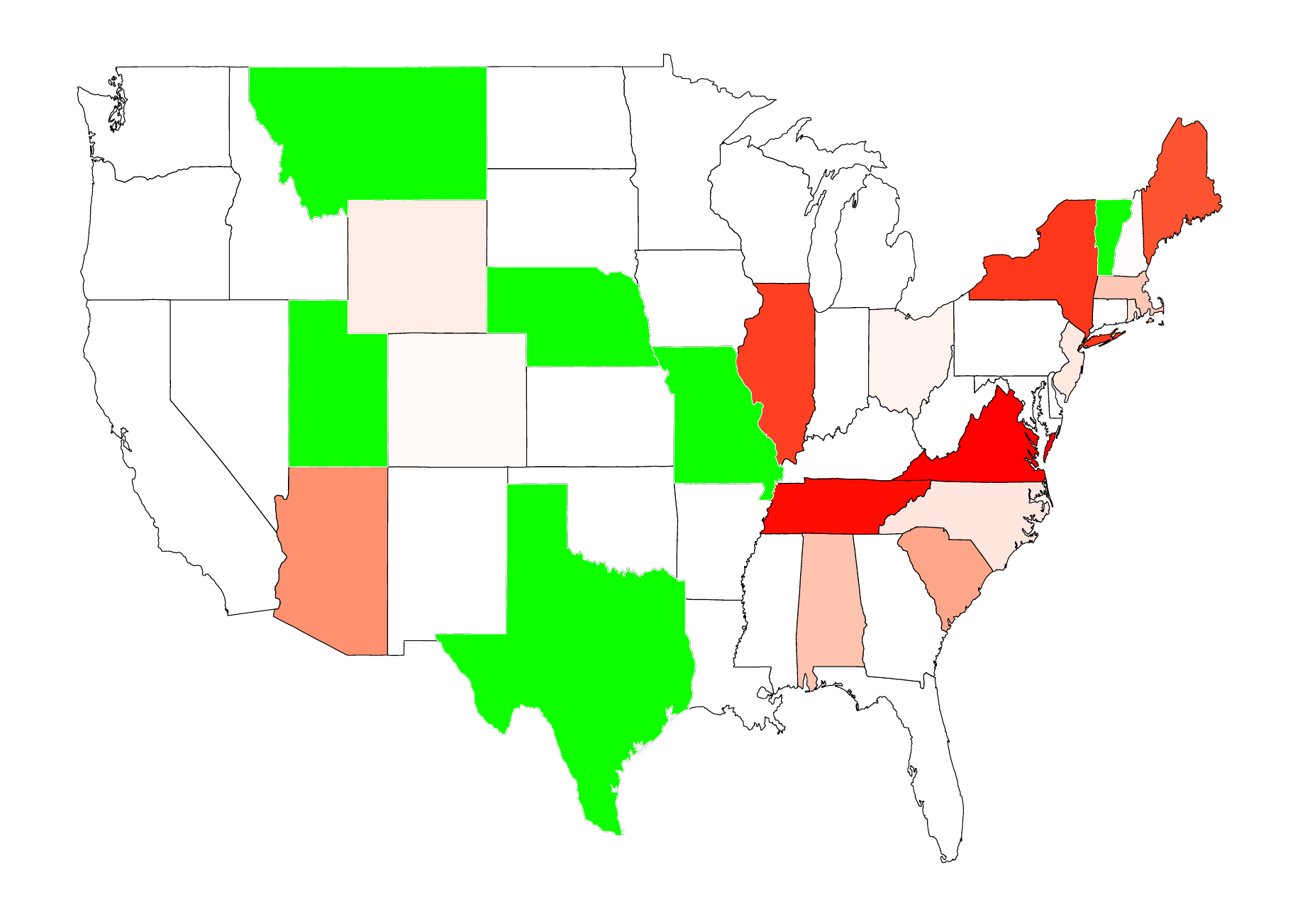} \\
		(d) SSD in category1  &
		(e) SSD in category2  &
		(f) SSD in category3  \\
		\includegraphics[width=0.31\textwidth]
		{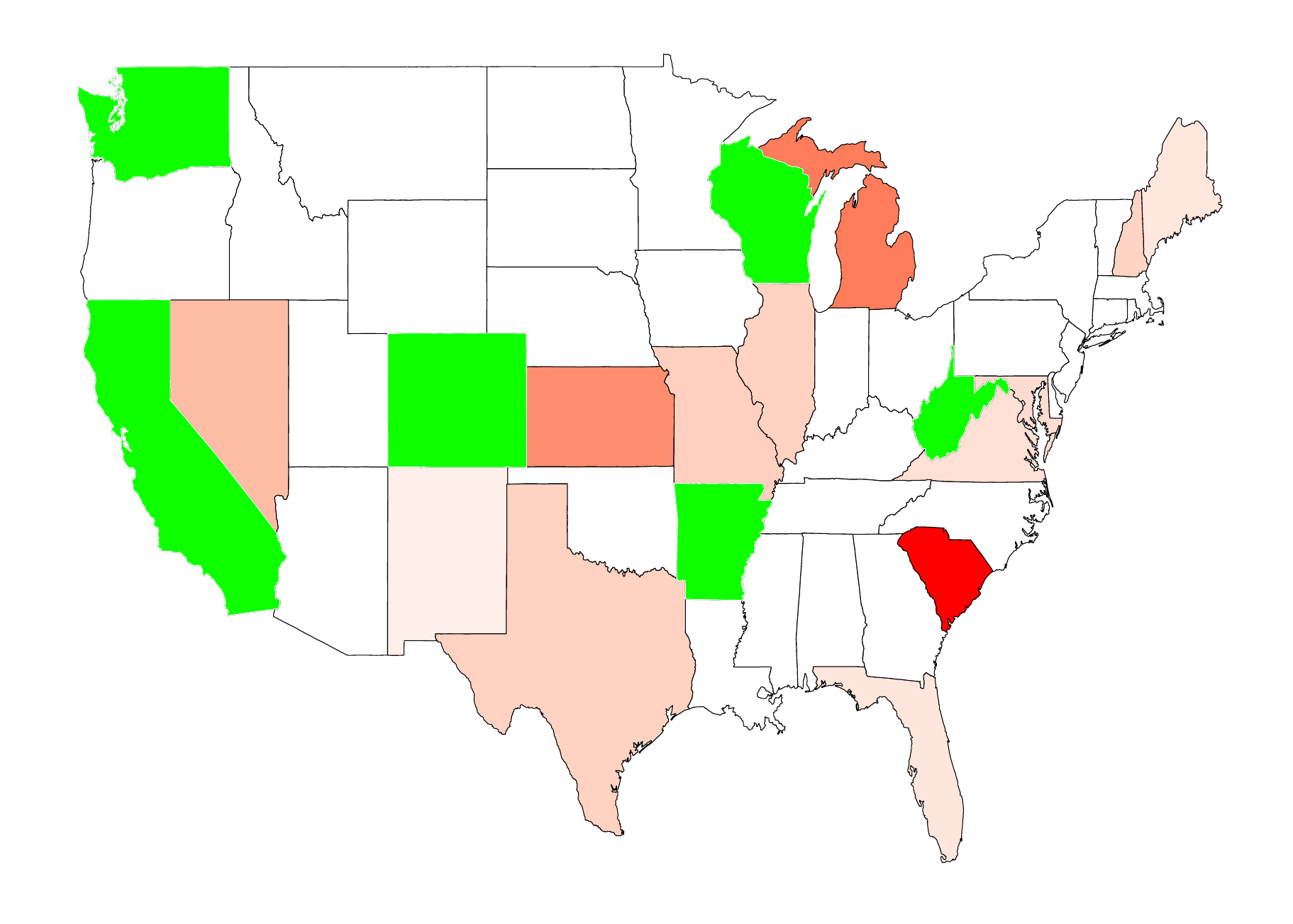}  &
		\includegraphics[width=0.31\textwidth]
		{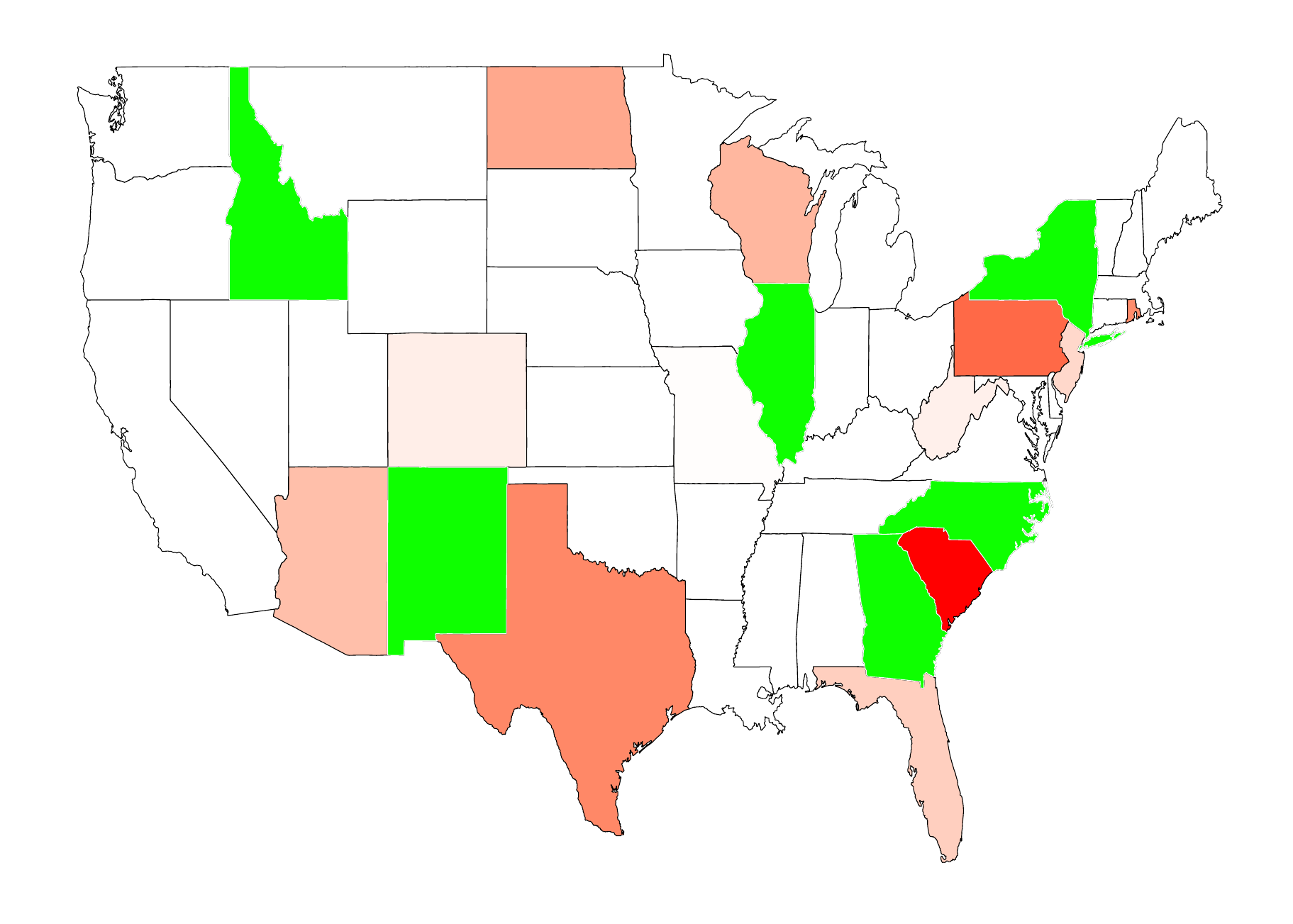}  &
		\includegraphics[width=0.31\textwidth]
		{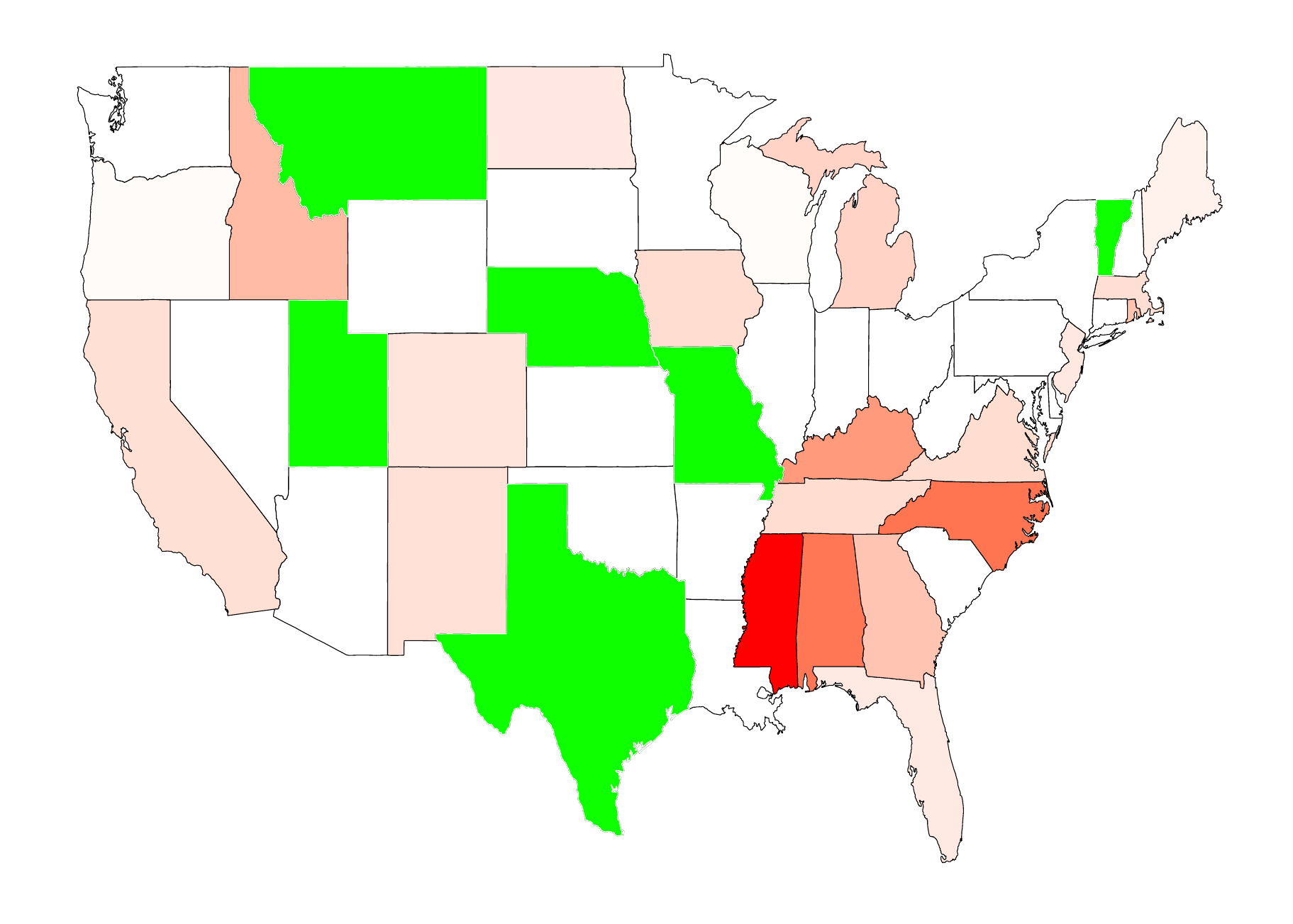} \\
		(g) SSR-Tensor in category1  &
		(h) SSR-Tensor in category2  &
		(i)  SSR-Tensor in category3 \\
	\end{tabular}
	\caption{Hot-spot detection performance by ZQ LASSO, SSD and our proposed SSR-Tensor method in Scenario 2 (decreasing global trend mean) with a large hot-spot of $\delta=0.5$. Here red is for the falsely detected hot-spots by the algorithm (i.e., false positive), blue refers to the undetected but true hot-spots (i.e., false negative), and green means the detected and true hot-spots (i.e., true positive))
	\label{fig: sim_hotspot_dect_map}  }
\end{figure}

    In addition, we also visualize our hot-spot detection results in Figure \ref{fig: sim_hotspot_dect_map}.
    From Figure \ref{fig: sim_hotspot_dect_map}, we can see that our proposed SSR-Tensor method can accurately detect the hot-spot location with the smallest false positive (i.e., red).

\begin{figure}[htbp]
	\begin{tabular}{cc}
		\centering
		\includegraphics[width=0.45\textwidth]{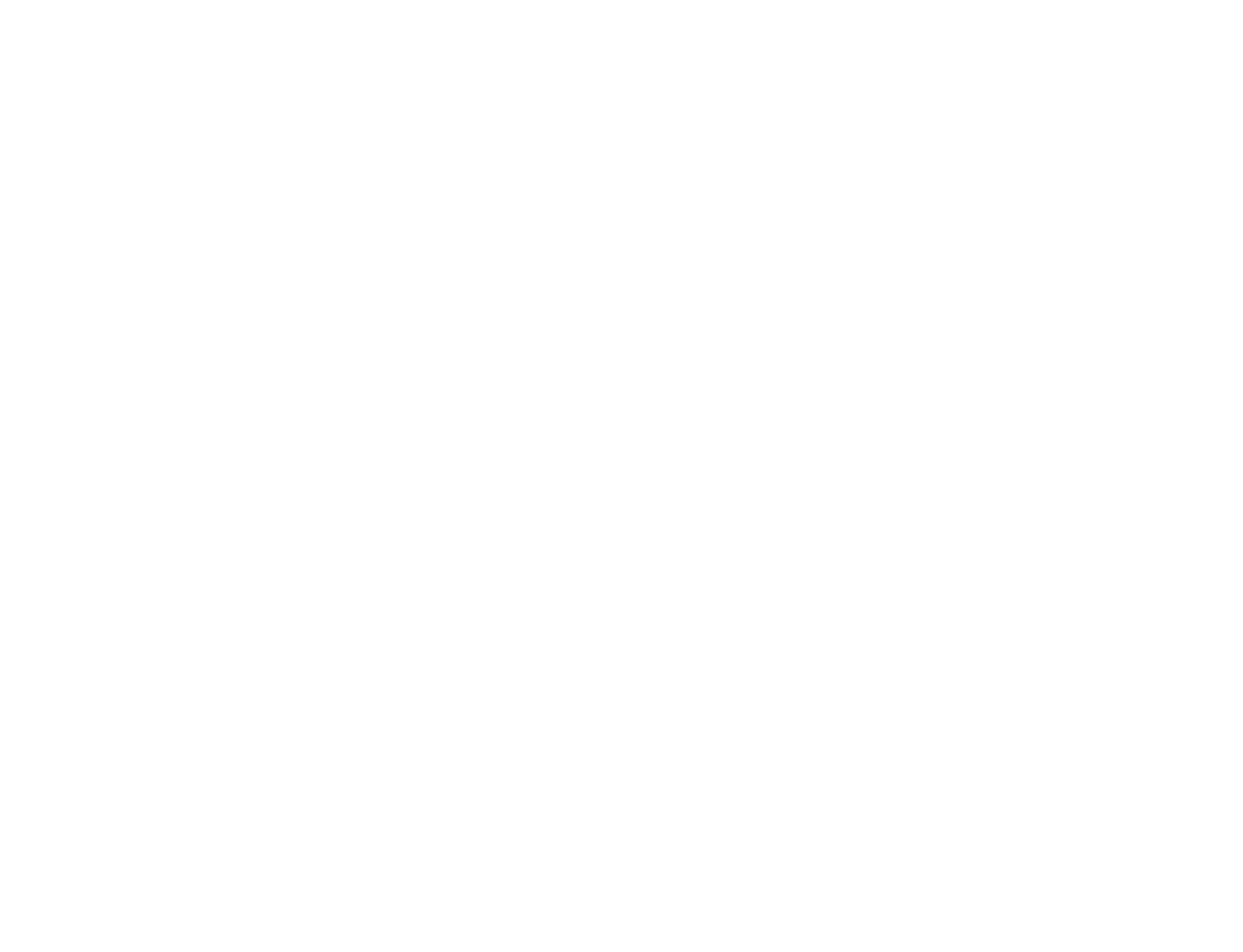}  &
        \includegraphics[width=0.45\textwidth]{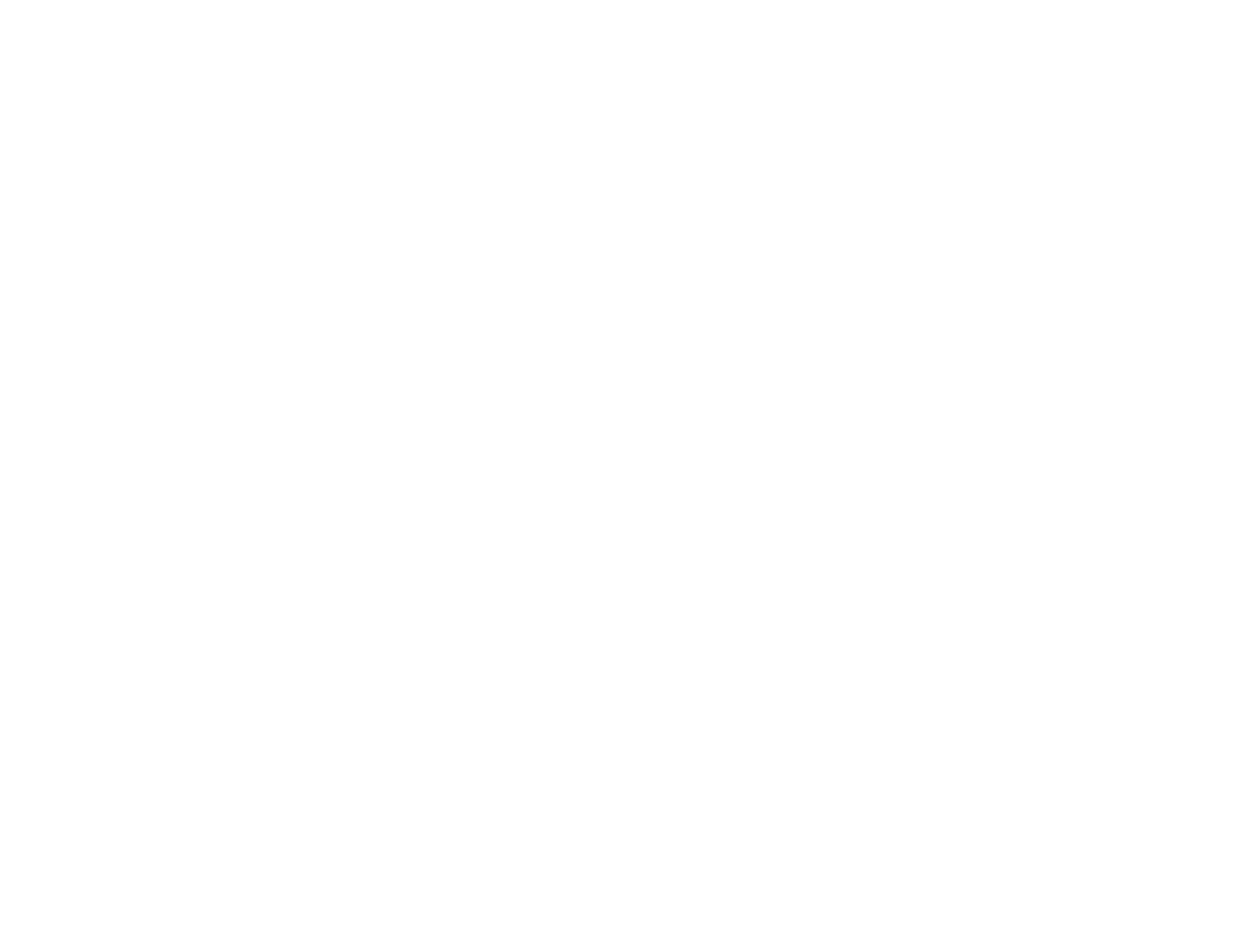} \\
		(a) ARL under no global trend mean &
		(b) ARL under decreasing global trend mean
	\end{tabular}
	\caption{ $\mbox{ARL}_1$ plot under different magnitude $\delta$ of the hot-spot
	\label{fig: sim_ARL_trend} 
}
\end{figure}

    Figure \ref{fig: sim_ARL_trend} illustrates the trend of the detection delay, $\rm{ARL}_1,$ of all methods, as $\delta$ changes from $0.1$ to $0.5$ with the step size of $0.1$.
    From the plot, our proposed SSR-Tensor method (the red curve) has the smaller detection delays than other baseline methods, particularly when there is a decreasing global trend mean and the magnitude of the hot-spot is small.
    Also, it is interesting to note that the detection delays of all methods are decreasing as the magnitude of the hot-spot is increasing, which is consistent with our intuition that it is easier to detect larger changes.
    In addition, note that the PCA method has the largest detection delays since it fails to consider the spatio-temporal correlation and the sparsity of the hot-spots.

\subsection{Background Fitness}
\label{sec: sim_background fitness}
    In this subsection, we will illustrate that our proposed SSR-tensor has a good estimation for the global trend  mean.
    To do this, we compare the Squared-Root of Mean Square Error (SMSE) of the fitness of a global trend mean in Table \ref{table: sim_background fitness SMSE}.
    Here we only compare our proposed method with SSD, since other baseline methods (ZQ LASSO, PCA, and T2) can not model the global trend mean.
    It is clear from Tables \ref{table: sim_background fitness SMSE} that our proposed SSR-tensor method does better in terms of the background fitness, especially in the scenario 2 (decreasing global trend mean).
\begin{table}[htbp]
    \centering
    \begin{tabular}{c|ccccc}
    \hline
    methods & $\delta=0.1$ & $\delta=0.2$ &$\delta=0.3$ & $\delta=0.4$ & $\delta=0.5$ \\
    \cline{1-6}
    & \multicolumn{5}{c}{ Scenario 1 (stationary global trend mean)}  \\
    \cline{2-6}
    SSR-tensor
    & \bf{0.0279}  &\bf{0.1712}  & \bf{0.1778} & \bf{0.1873} & \bf{0.1997}\\
    &(0.0024)      &(0.0025)     &(0.0025)     &(0.0024)     &(0.0020)\\
    SSD
    & 0.1779  &0.1826  & 0.1928 & 0.2076 & 0.2098\\
    &(0.0101) &(0.0052)&(0.0029)&(0.0024)&(0.0023)\\
    \cline{1-6}
    & \multicolumn{5}{c}{  Scenario 2 (decreasing global trend mean) }  \\
    \cline{2-6}
    SSR-tensor
    &\bf{0.0030}   & \bf{0.0030} & \bf{0.0030} & \bf{0.0031} & \bf{0.0031}\\
    &(0.0072)      &(0.0068)     &(0.0061)     &(0.0064)     &(0.0060)\\
    SSD
    &0.2778   & 0.2725 &0.2781  & 0.2842 &0.2602\\
    &(0.0136) &(0.0114)&(0.0143)&(0.0113)&(0.0084)\\
    \hline
    \end{tabular}
\caption{SMSE in two scenarios with different shift
\label{table: sim_background fitness SMSE}
}
\end{table}


\section{Case Study}
\label{sec: case study}
    In this section, we apply our proposed SSR-tensor method to the crime rate dataset described in  Section \ref{sec: data}.
    Our proposed method is compared with other benchmarks (the same benchmarks we used in Section \ref{sec: simulation}) from two aspects,
    one is the performance in the temporal detection of  hot-spots (i.e., which year it occurs)
    and the other is the performance in the localization of the hot-spots (i.e., which state and which type of crime rates may trigger the alarm).

\begin{figure}[htbp]
	    \centering
        \includegraphics[width=0.75\textwidth]{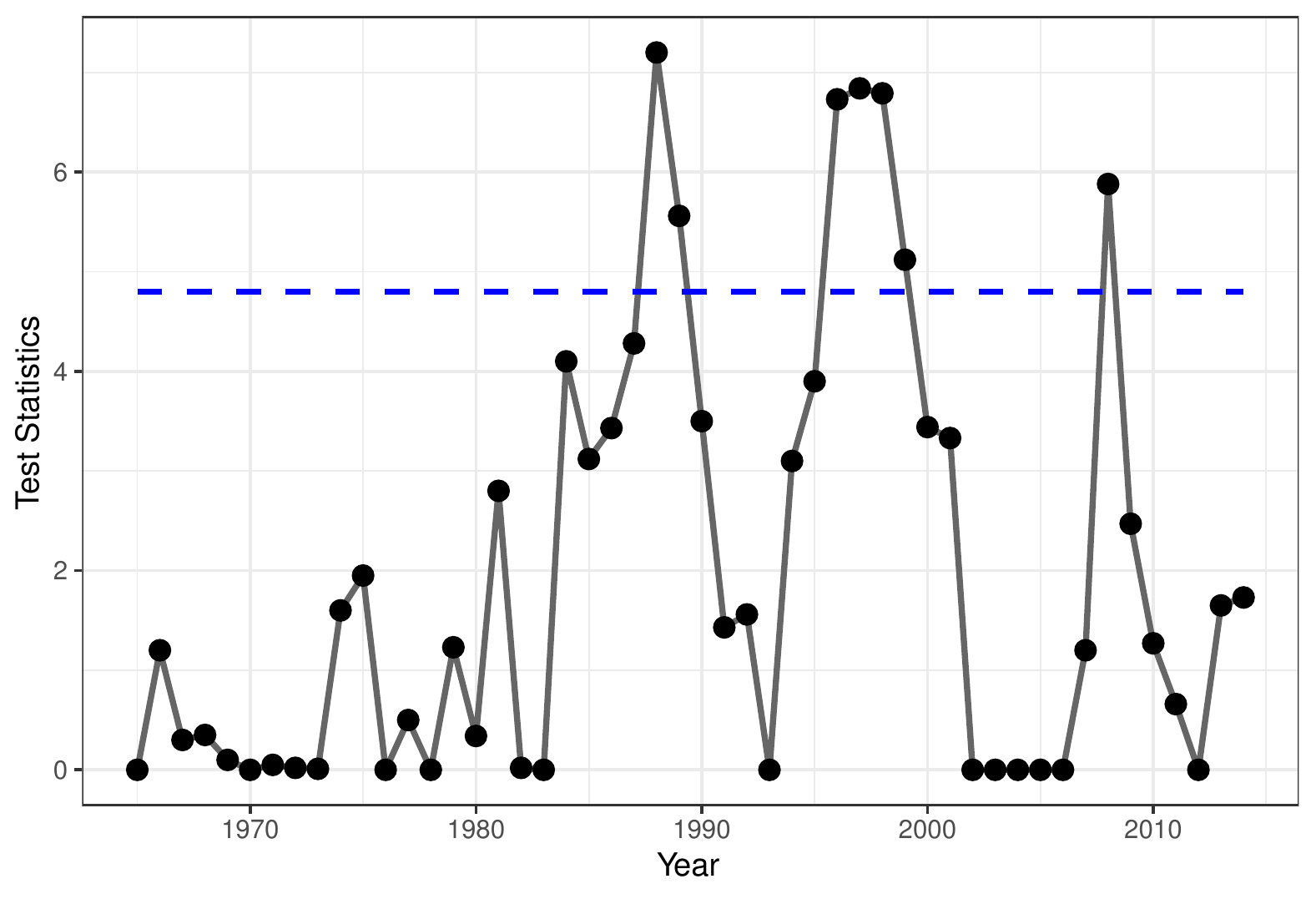}
	    \caption{Detection of hot-spot  by SSR-Tensor
	    \label{fig:control chart CUSUM} }
    \end{figure}

    First, we compare the performance of the detection delay of the hot-spot.
    For our proposed SSR-Tensor method, we build a CUSUM control chart utilizing the test statistic in Section \ref{sec: Temporal Detection and Spatial Location of Hot Spot}, which is shown in Figure \ref{fig:control chart CUSUM}.
    From this plot, we can see that the hot-spots are detected at $24-25$-th, $32-35$-th  and $44$-th years, i.e, 1989-1990, 1997-2000 and 2009.

    For the benchmark methods for comparison, we also apply
    SSD \cite[see][]{SSD},
    ZQ LASSO \cite[see][]{zou2009multivariate},
    PCA \cite[see][]{PCA} and
    T2 \cite[see][]{T2}
    to the crime rate dataset and summarize the performance of the detection of a hot-spot in Table \ref{table: case study temporal detection}.
    Note that all the temporal changes reported in Table \ref{table: case study temporal detection} is the  year with the first alarm.
    Clearly, our proposed SSR-tensor method achieves the fastest detection of  the hot-spot compared to all other benchmark methods.
    Numerically verified  in simulation, we can say that it is of quite likely that 1989-1990, 1997-2000 and 2009 are indeed hot-spots, and our proposed SSR-tensor method works well in the crime rate dataset.

    \begin{table}[htbp]
        \centering
        \scriptsize
        \begin{tabular}{c|ccccc}
        \hline
        methods & SSR-Tensor & SSD & ZQ LASSO & PCA & T2 \\
        \cline{2-5}
        \hline
        NO. of year that the first temporal changes & \bf{24} & 30 & None & None & None\\
        \hline
        \end{tabular}
        \caption{Detection of hot-spot in crime rate dataset
        \label{table: case study temporal detection}
        }
    \end{table}

\begin{figure}[htbp]
    \centering
	\begin{tabular}{ccc}
		\includegraphics[width=0.31\textwidth]{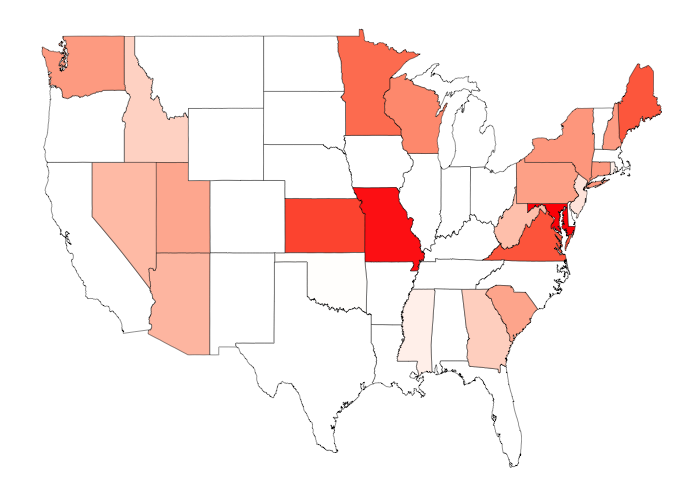}  &
		\includegraphics[width=0.31\textwidth]{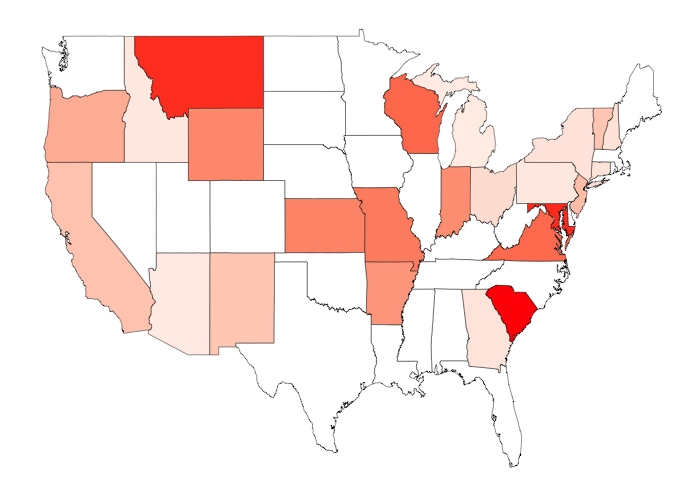}  &
		\includegraphics[width=0.31\textwidth]{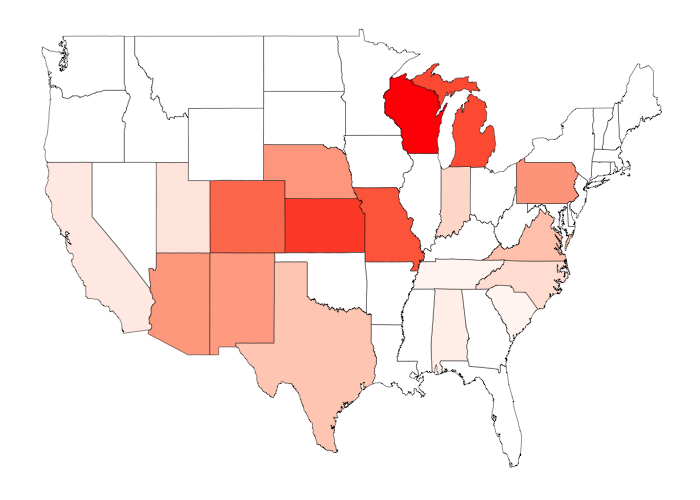}  \\
		(a)  1989, $r_1$, SSR-tensor  & (b) 1989, $r_2$ , SSR-tensor & (c) 1999, $r_3$, SSR-tensor
		\tabularnewline
		\includegraphics[width=0.31\textwidth]{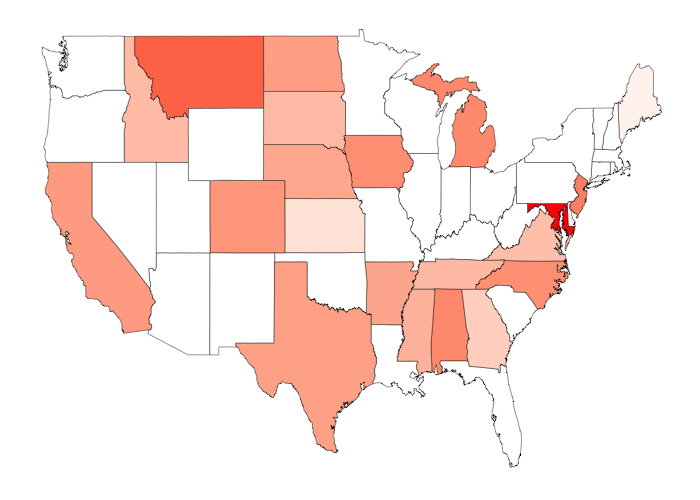}  &
		\includegraphics[width=0.31\textwidth]{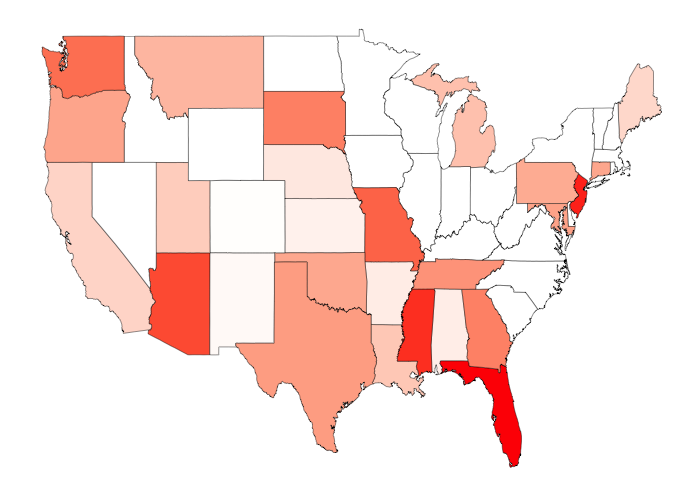}  &
		\includegraphics[width=0.31\textwidth]{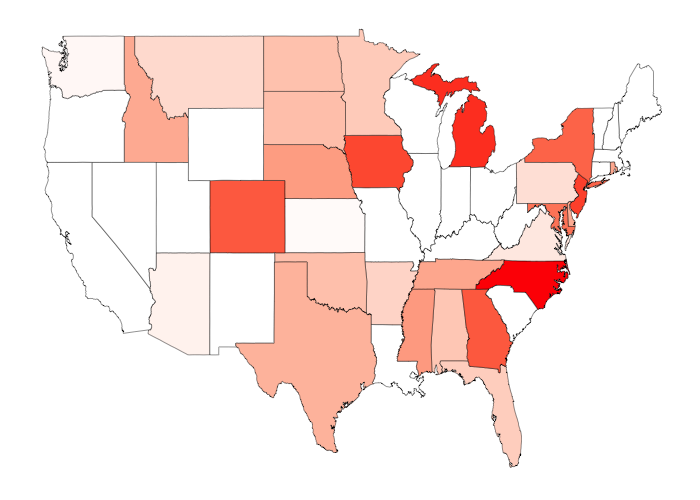}  \\
		(d)  1995, $r_1$, SSD  & (e) 1995, $r_2$, SSD & (f) 1995, $r_3$, SSD
	\end{tabular}
	\caption{Localization of hot-spot. Top row: three crime rates by our proposed SSR-Tensor method. Bottom row: three crime rates by the SSD.
             The red color of the states means that there is an upward shift for this corresponding state and the deeper the color, the larger the increase of the rate.
	\label{fig: hot-spot map} }
    \end{figure}

    Next, after the detection of  hot-spots, we need to further localize the hot-spots in the sense that we need to find out which state and which type of crime rate may lead to the occurrence of a hot-spot.
    Because the baseline methods, PCA and T2, can only detect when the changes happen and ZQ-LASSO fails to detect any  changes, we only show the localization of hot-spot by SSR-Tensor and SSD, where the results are visualized in Figure \ref{fig: hot-spot map}.

\begin{figure}[htbp]
    \centering
	\begin{tabular}{cc}
		\includegraphics[width=0.45\textwidth]{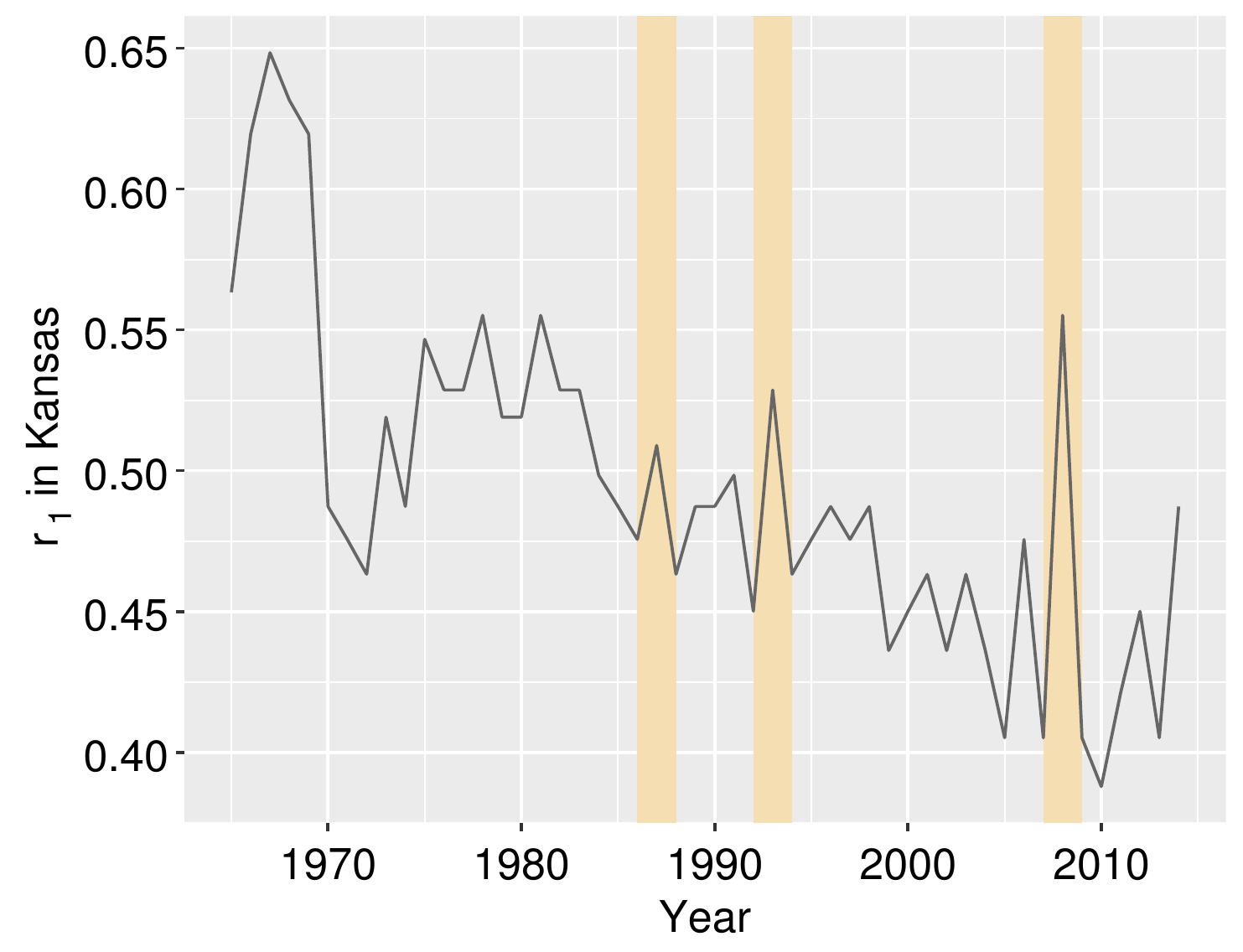}  &
		\includegraphics[width=0.45\textwidth]{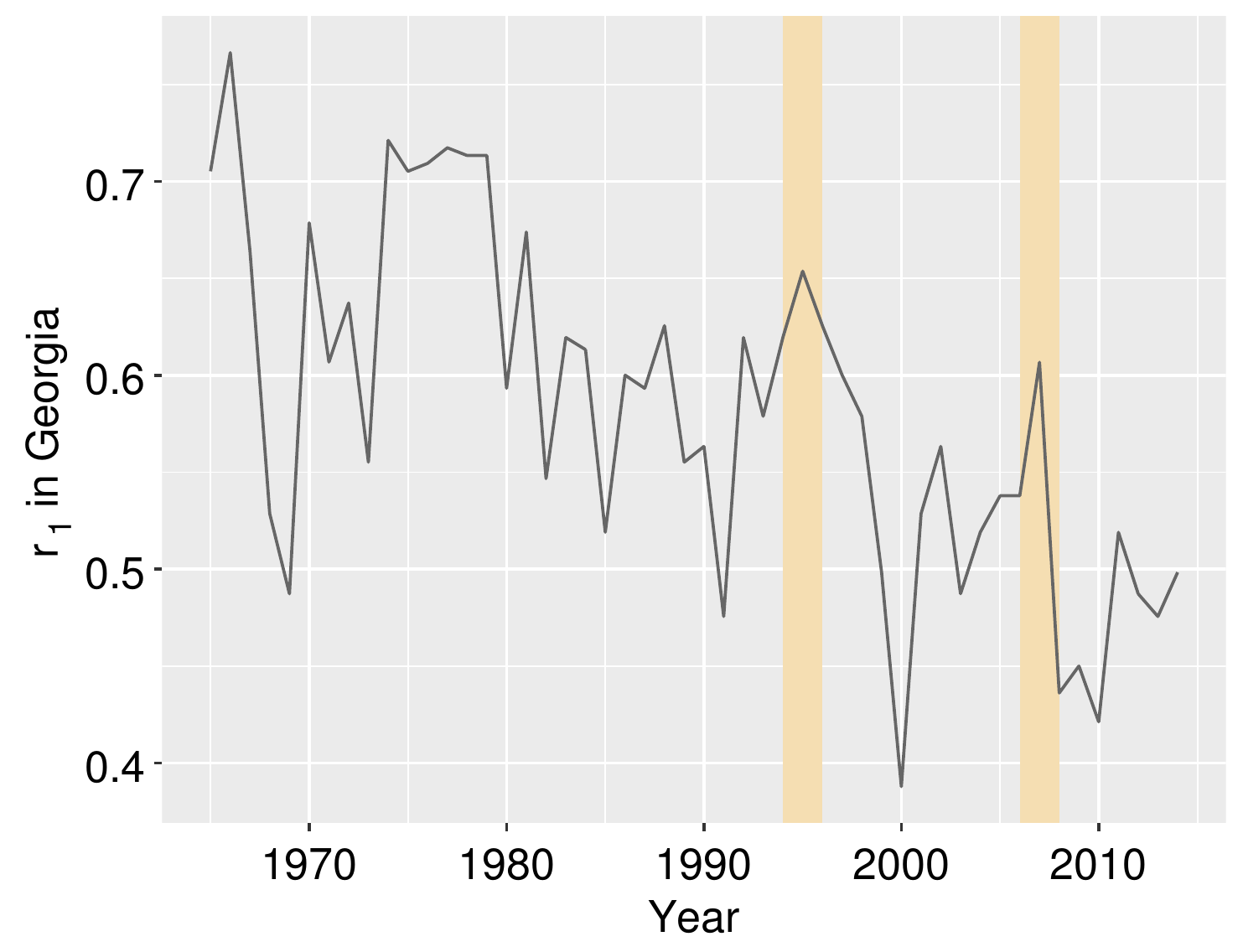}  \\
		(a)  Kansas & (b) Georgia
	\end{tabular}
	\caption{Time series plot of the first type of crime rate in Kansas(a) and Georgia(b).
	\label{fig: case study explain the hot-spot}  }
    \end{figure}

    Compared to SSD, it can be seen that our proposed SSR-tensor method detects some localized sparse hot-spots, which might be useful to identify where the sudden increase of crime happens.
    As an example, let us consider Kansas, which is declared as a hot-spot in all the three temporal changes year 1989, 1997 and 2009.
    Figure  \ref{fig: case study explain the hot-spot} shows the time series plot of the first type of crime rate in Kansas.
    The plot shows that, during the decreasing trend of the last 30 years, Kansas experienced some sudden increase, which may cause it to be detected as a hot-spot by SSR-Tensor.
    Another example is Georgia, which is declared as a hot-spot in $1995$.
    One possible explanation is that Gerogia experience huge floods caused by hurricane and tropical storm.
    The floods tended to damage people's houses and cut their food supply, which possibly leads to an increase in crime rates.
    So we can see from Figure \ref{fig: case study explain the hot-spot}(b) that, during the global decreasing trend, Georgia experienced an sudden increase around 1995.
    However, without our SSR-Tensor model, it may be difficult to tell whether it is a global trend or a hot-spots.

\section{Conclusion}
    Hot-spot detection in spatio-temporal data is an important problem in real life.
    In this paper, we propose the `SSR-Tensor' method for the detection of hot-spots in spatio-temporal data.
    Unlike existing methods for hot-spot detection, which is only workable for scalar or functional data, our proposed SSR-Tensor method is able to decompose the variate pattern of the multi-dimension data  into the global trend mean, local hot-spots and residuals.
    The estimation of hot-spot is solved by optimizing the sum of residuals with two penalty terms, which controls the sparsity of the hot-spots and the temporal consistency of the hot-spots, respectively.

    To efficiently solve the above-mentioned high-dimensional optimization problems,  we first reduce the unknown parameters and then simplify the generalized LASSO problem into the regular LASSO problem where many well-known LASSO algorithms can be used.
    We chose the FISTA algorithm in our paper because it currently has the fastest convergence rate up.
    We compare our proposed SSR-Tensor method with other benchmarks in terms of detection accuracy, computational time and background fitness.
    Based on Monte Carlo simulations and the case study of the crime rate dataset, we conclude that overall our proposed SSR-Tensor method outperforms other benchmarks.
    While the classical statistical process control (SPC) or sequential change-point detection problems have been studied for several decades, research on the hot-spot of tensor data is rather limited, mainly due to computational complexity and sparsity of hot-spots.
    Clearly, there are many opportunities to improve the algorithms and methodologies.
    For instance, it will be interesting to study the relationship between our proposed SSR-Tensor method and the autoregressive model.
    Moreover, in this paper, we specify the tensor basis, and then investigate its performance on the detection and localization of the hot-spot. It will be useful to investigate the robustness  effects of different tensor bases.
    Finally, further research direction includes hot-spot detection problems with more complicated spatiotemporal dynamics.

\bibliographystyle{tfs}
\bibliography{myreference_SSRtensor}

\appendix
\section{Proof of Fast Calculation of $\mathbf{y}^{*}$ via Tensor Algebra}
\label{proof: prop 1}
\begin{proof}
\begin{eqnarray*}
\mathbf{y}^{*}
&= &
[\mathbf{I} - \mathbf{B}_{m} (\mathbf{B}_{m}' \mathbf{B}_{m})^{-1} \mathbf{B}_{m}'] \mathbf{y}\\
&= &
\mathbf{y} - \mathbf{B}_{m}( ( \mathbf{B}_{m,s}'\otimes \mathbf{B}_{m,r}'\otimes \mathbf{B}_{m,t}')(\mathbf{B}_{m,s} \otimes \mathbf{B}_{m,r} \otimes \mathbf{B}_{m,t}))^{-1})\mathbf{B}_{m}'\mathbf{y}\\
&= &
\mathbf{y} - \mathbf{B}_{m}((\mathbf{B}_{m,s}' \mathbf{B}_{m,s}) \otimes(\mathbf{B}_{m,r}'\mathbf{B}_{m,r})\otimes(\mathbf{B}_{m,t}'\mathbf{B}_{m,t}))^{-1})\mathbf{B}_{m}'\mathbf{y}\\
&= &
\mathbf{y} - \mathbf{B}_{m}((\mathbf{B}_{m,s}' \mathbf{B}_{m,s})^{-1} \otimes( \mathbf{B}_{m,r}' \mathbf{B}_{m,r})^{-1} \otimes( \mathbf{B}_{m,t}' \mathbf{B}_{m,t})^{-1})\mathbf{B}_{m}' \mathbf{y}\\
&= &
\mathbf{y} - (\mathbf{B}_{m,s}(\mathbf{B}_{m,s}' \mathbf{B}_{m,s})^{-1} \mathbf{B}_{m,s}') \otimes( \mathbf{B}_{m,r}( \mathbf{B}_{m,r}' \mathbf{B}_{m,r})^{-1} \mathbf{B}_{m,r}') \\
& &
\otimes( \mathbf{B}_{m,t}( \mathbf{B}_{m,t}'\mathbf{B}_{m,t})^{-1} \mathbf{B}_{m,t}') \mathbf{y}\\
&= &
\mathbf{y} -
\mathcal{Y}
\times_1 ( \mathbf{B}_{m,s}( \mathbf{B}_{m,s}' \mathbf{B}_{m,s})^{-1} \mathbf{B}_{m,s}')
\times_2 ( \mathbf{B}_{m,r}(\mathbf{B}_{m,r}' \mathbf{B}_{m,r})^{-1} \mathbf{B}_{m,r}')\\
& &
\times_3 (\mathbf{B}_{m,t}(\mathbf{B}_{m,t}' \mathbf{B}_{m,t})^{-1}\mathbf{B}_{m,t}')
\end{eqnarray*}
\end{proof}

\section{Proof of Proposition \ref{prop: generalized lasso to lasso}}
\label{proof: prop 2}
\begin{proof}
	By introducing matrix $\widetilde{\mathbf D},$ we change variables to $\widehat{\boldsymbol\beta}=(\widehat{\boldsymbol\beta}_1,\widehat{\boldsymbol\beta}_2)=\widetilde{\mathbf D}\boldsymbol\theta_{h,0,\lambda_{2}}.$
	Accordingly, $\left\Vert \mathbf D\boldsymbol\theta_{h,0,\lambda_{2}}\right\Vert _{1}=\left\Vert \boldsymbol\beta_{1}\right\Vert _{1}$
	because the rows in matrix $A$ are orthogonal to those in matrix
	$\mathbf D$.
	Besides,
    $
    \mathbf X\boldsymbol\theta_{h,0,\lambda_{2}}
    =
    \mathbf X\widetilde{\mathbf D}^{-1}\boldsymbol\beta
    =
    \mathbf{X}_{1} \boldsymbol\beta_{1} + \mathbf{X}_{2} \boldsymbol\beta_{2}
    $.
	Therefore, we can rewrite equation \eqref{equ: estimation2} as.
	\begin{equation}
	\arg\min_{\boldsymbol{\theta}_{h,\lambda_{1},\lambda_{2}}}
    \Vert\mathbf{y}^{*}-(\mathbf X_{1}\boldsymbol\beta_{1}+\mathbf X_{2}\boldsymbol\beta_{2})\Vert_2^2
    +
    \lambda_{2}\Vert\boldsymbol\beta_{1}\Vert_{1}
	\label{equ: genlasso rewrite}
	\end{equation}
	
	By taking derivative to $\boldsymbol\beta_{2},$we have
	\begin{equation}
	\boldsymbol\beta_{2}=(\mathbf X_{2}'\mathbf X_{2})^{-1}\mathbf X_{2}'(\mathbf y - \mathbf X_{1} \boldsymbol\beta_{1}).
	\label{equ: genlasso diff to beta2}
	\end{equation}
	
	By plugging  in equation \eqref{equ: genlasso diff to beta2}, we can
	rewrite equation \eqref{equ: genlasso rewrite} as that shown in equation
	\eqref{equ: generalized lasso to lasso}. There are lots of R package
	available for solving lasso, such as \textit{LARS} and \textit{glmnet}, from which we
	can back-transform to get the generalized lasso solution $\widehat{\boldsymbol\theta}_{h,0,\lambda_{2}}=\widetilde{\mathbf D}^{-1}\widehat{\boldsymbol\beta}$.
	
\end{proof}

\end{document}